\documentclass[aps,prd,reprint,nofootinbib,longbibliography, amsmath,amssymb]{revtex4-1}
\usepackage{blindtext}
\usepackage{mathtools}
\usepackage{cancel}
\usepackage[usenames,dvipsnames]{color}
\usepackage{adjustbox}
\usepackage{graphicx}
\usepackage[ISO]{diffcoeff}[=v4]
\usepackage{bm}% bold math
\normalsize
\usepackage{booktabs}
\usepackage{siunitx}
\usepackage{tabularx}
\newcolumntype{Y}{>{\centering\arraybackslash}X}
\usepackage{soul}
\usepackage[utf8]{inputenc}
\usepackage[english]{babel}
\usepackage{hyperref}
\hypersetup{
    colorlinks=true,
    linkcolor=blue,
    citecolor=red,
    filecolor=magenta,      
    urlcolor=magenta,
}
\usepackage[capitalise]{cleveref}
\def\sfrac#1#2{\ensuremath{\textstyle \frac{#1}{#2}}}

\newcommand\funop[1]{\mathop{{}#1}}
\newcommand{\usim}{\mathord{\sim}}
\renewcommand{\vec}[1]{\bm{#1}}

\begin{document}
\title{Rescuing the primordial black holes all-dark matter hypothesis\\
from the fast radio bursts tension}

\author{Dorian W.~P.~Amaral$^{1\dagger}$ and Enrico D.~Schiappacasse$^{2*}$}
\affiliation{
$^1$Department of Physics and Astronomy, Rice University, Houston, Texas 77005, USA\\
$^2$Facultad de Ingenier\'ia, Arquitectura y Diseño, Universidad San Sebasti\'an,
Santiago, Chile\\
}

\begin{abstract}
\noindent
Primordial black holes (PBHs) as an all-dark matter (DM) hypothesis has recently been demotivated by the prediction that these objects would source an excessive rate of fast radio bursts (FRBs). However, these predictions were based on several simplifying assumptions to which this rate is highly sensitive. In this article, we improve previous estimates of this rate arising from the capture of PBHs by neutron stars (NSs), aiming to revitalise this theory. We more accurately compute the velocity distribution functions of PBHs and NSs and also consider an enhancement in the NS and DM density profiles at galactic centers due to the presence of a central supermassive black hole. We find that previous estimates of the rate of FRBs sourced by the capture of PBHs by NSs  were 3 orders of magnitude too large, concluding that the PBHs as all DM hypothesis remains a viable theory and that the observed FRB rate can only be entirely explained when considering a central, sufficiently spiky PBH density profile.
\end{abstract}
\maketitle

%%%%%%%%%%%%%%%%%%%%%%%%%%%%%%%%%%%%%%%%%
\section{Introduction}
%%%%%%%%%%%%%%%%%%%%%%%%%%%%%%%%%%%%%%%%%

The nature of dark matter (DM) is one of the largest outstanding puzzles in modern physics. While it has
been confirmed by astrophysical and cosmological observations,
its fundamental properties remain unclear. Excepting the fact that it must interact gravitationally, its other properties---such as its mass, spin, and other potential interactions---remain a mystery~\cite{Peebles:2013hla, 
Freese:2017idy, Bertone:2016nfn}. 
In the dark sector, 
the weakly interacting massive particle (WIMP)~\cite{Jungman:1995df} and the axion~\cite{PRESKILL1983127, ABBOTT1983133, DINE1983137, Kim:2008hd} constitute two popular DM candidates, but they remain elusive despite three decades of collider\,\cite{dEnterria:2021ljz, Buchmueller:2017qhf} and direct DM searches\,\cite{2014PDU.....4...14R,Roszkowski:2017nbc, Schumann:2019eaa}.  
Another class of DM candidates is the primordial black hole (PBH)~\cite{Hawking:1971ei, 1975Natur.253..251C}. Being stable and cold on cosmological scales, PBHs make for an outstanding DM candidate. 
While most of the PBH parameter space has been ruled out by dynamical and relic observations, they  can still account for the entirety of the DM content in the Universe if their masses lie in the range $10^{-14}\, M_{\odot}\text{--}10^{-11}\,M_{\odot}$~\cite{Inomata:2017okj} due to uncertainties in certain bounds, such as those arising from microlensing and white dwarfs~\cite{PhysRevLett.80.1138, Takahashi:2003ix, Carr:2020gox}.
Indeed, if PBHs make up all of DM, a \textit{striking consequence} is that beyond Standard Model interactions are not needed to explain the DM we currently observe in the Universe\footnote{However, we note that, for example, beyond Standard Model inflationary theories are required to understand the production of PBHs in the early universe in the abundance required to make up all of DM (see, for instance, ~\cite{Inomata:2018cht}).}. Therefore, from the viewpoint of high-energy physics, the PBH all-DM scenario is a highly compelling theory.
However, the PBH all-DM hypothesis has recently been jeopardized. It has been predicted that
these astrophysical objects would source a rate of fast radio bursts (FRBs)\,\cite{Fuller:2017uyd, Katz:2018xiu} exceeding that inferred from observations\,\cite{Abramowicz_2018}. Fast radio bursts are bright and brief radio frequency pulses whose origin remains unknown in spite of the great physics community  effort. 
Among the several proposed mechanisms for producing FRBs, neutron stars (NSs) 
are a popular possibility due to their large gravitational, magnetostatic, and electromagnetic energy densities \,\cite{Totani:2013lia, Lipunov:2013axa, Yamasaki:2017hdr,  DasGupta:2017uac,  2017ApJ...838L...7D, Waxman:2017zme, Zhang:2018csb, Buckley:2020fmh}.
The key idea is that encounters between NSs and PBHs in galaxies would lead to the capture of the PBH by the dense neutron star medium. The PBH would then swallow the host, resulting in a short, bright energy burst during the final stage of the NS collapse. Assuming a uniform distribution for the observed FRBs among $10^{11}$ galaxies, the authors of Ref.~\cite{Abramowicz_2018} predicted an FRB rate 3--4 orders of magnitude larger than the observed rate should the entirety of DM be composed of PBHs. Subsequently, based on the calculation of the PBH capture rate by NSs during close encounters performed in Ref.~\cite{Genolini:2020ejw}, one of the present authors and others in Ref.~\cite{Kainulainen:2021rbg} concluded that the presumed tension between FRB observations and the PBH all-DM hypothesis is just an artifact of a PBH capture rate overestimation. With the goal of comparison with Ref.~\cite{Abramowicz_2018}, authors in  Ref.~\cite{Kainulainen:2021rbg} used the same simplified toy model for the galaxy and velocity distributions. However, the rate of FRBs is highly sensitive to these assumptions, and more realistic modeling is needed for a robust estimate of it.
In this article,  we make significant improvements to the previous estimates performed in Ref.~\cite{Kainulainen:2021rbg} of the FRB rate arising from the PBH capture by NSs under the PBH all-DM hypothesis. In such a scenario, since the production of FRBs is closely linked to the galactic NS and DM density profiles and NS/PBH relative velocity distribution, we mainly focus on these features. First, as a benchmark model for a spiral galaxy, we consider the Milky Way model obtained in\, Ref.~\cite{Bhattacharjee:2012xm} (the BCKM-model), which is based on the most likely values of the relevant galactic parameters from a Markov Chain Monte Carlo analysis using data from  galactic rotation curves. Second, we use the inversion Eddington formula and a series of Monte Carlo (MC) simulations to compute the relative velocity distribution function (RVDF) between PBHs and NSs in the outer regions of the galaxy and, for the inner regions, we use the Jeans equation.
In the original works on FRBs in the context of the PBH all-DM hypothesis~\cite{Abramowicz_2018, Kainulainen:2021rbg}, a galactic toy model for a spiral galaxy composed of a bulge and disk was used to predict the total FRB rate. To go further in this framework and provide a more representative picture of the FRB rate sourced by NS-PBH captures,  we must, for example, consider the effects of matter profiles and velocity distributions. In this work, we have improved upon the predictions of this framework, showing that our modeling improvements lead to sizeable effects on the FRB predictions in the context of the PBH all-DM hypothesis. We have assumed that spiral galaxies dominate the FRB signal and used the Milky Way as a representative spiral. A complete treatment of the FRB rate sourced by NS-PBH captures would require detailed knowledge of the galactic populations surrounding the Milky Way, including each of their presumed DM and NS profiles. We do not attempt to do this here\textcolor{black}{; 
however, we briefly comment on the assumption made in the original work\,\cite{Abramowicz_2018} and a possible extension of this article in the \cref{Sec:RD}.}
Our results qualitatively confirm the previous estimates performed in\, Ref.~\cite{Kainulainen:2021rbg} and further release the tension between FRBs and the PBH all-DM hypothesis reported in Ref.~\cite{Abramowicz_2018}. In addition, our model improvements with respect to previous estimates lead us to conclude that the production of FRBs via PBH-NS encounters is  $\usim 3$ orders of magnitude smaller than the observed FRB rate. The reason is that typical relative velocities between PBHs and NSs, more accurately computed via the method presented here, are too large to allow for the PBH capture by the NS. 
Since the predicted FRB rate is highly dependent on the baryonic and DM density profiles at galactic centers, we also  estimate such a rate considering density enhancements. For spiral galaxies, which dominate the signal, we find that if they host large enough spiky PBH central densities, the predicted FRB rate in the PBH all-DM hypothesis may be consistent for an order-of-magnitude estimate with the observed FRB rate.

%%%%%%%%%%%%%%%%%%%%%%%%%%%%%%%%%%%%%%%%%
\section{Galactic Model}
\label{sec:gal-model}
%%%%%%%%%%%%%%%%%%%%%%%%%%%%%%%%%%%%%%%%%

We use the BCKM-model\,\cite{Bhattacharjee:2012xm} for the Milky Way Galaxy as a benchmark model for a spiral galaxy. The density of the DM halo,  $\rho_{\text{DM}}$, follows a Navarro-French-White (NFW) profile\,\cite{1996ApJ...462..563N} normalized to the DM density at the solar position, $\rho_{\text{DM},\,\odot}$, according to\,\footnote{We point out that there is some debate on the nature of the DM profile close to the galactic center; the profile could instead be, for example, a cored or cuspy NFW profile (see, for instance, the discussion in Sec. III of \cite{Hertzberg:2020kpm}). This would lead to either a suppression or enhancement of the predicted FRB rate, respectively. In Sec.~\ref{sec:enhanced-density}, we study the effect of a central supermassive black hole on the DM profile at the Milky Way galactic center, including spiky and cuspy profiles.}
\begin{equation}
\rho_{\text{DM}}(r) = \rho_{\text{DM},\,\odot} \left( \frac{R_0}{r} \right)\left( \frac{r_s + R_0}{r_s + r} \right)^2\,,
\label{eq:rhodm}
\end{equation}
where $r \equiv |{\bf{x}}|$ is the galactic radius,  $r_s$ is the scale radius, and $R_0$ is the solar distance from the center of the Galaxy.  The visible matter is formed by a spheroidal bulge overlapped with an axisymmetric disk. The bulge decays with the galactic radius  as $r^{-3}$ for large distances and depends on the central density $\rho_{b,0}$ and
the scale radius $r_b$ of the bulge. The galactic disk shows a double exponential decay as we depart from the galactic plane and center, respectively. The distribution densities of the bulge ($\rho_b$) and disk ($\rho_d$) are 
\begin{align}
\rho_{b}(r) &= \rho_{b,0} \left[ 1 + \left(\frac{r}{r_b}\right)^2  \right]^{-3/2}\,,\label{eq:rhob}\\
\rho_d(R,\,z)&= \frac{\Sigma_{\odot}}{2z_d}\exp\left(-\frac{R-R_0}{R_d}\right)\exp\left(-\frac{|z|}{z_d}\right)\,,
\end{align}
where $z_d$ and $R_d$ are the height and radius scale lengths for the disk, respectively, $\Sigma_{\odot}$ is the local surface density at the solar position, and $r = (R^2 + z^2)^{1/2}$ in axisymmetric cylindrical coordinates. By performing a Markov Chain Monte Carlo analysis based on observed galactic rotation curve data, Ref.~\cite{Bhattacharjee:2012xm} found the most likely values for these parameters, which we summarize in Table\,\ref{tab:gal-model}.
\begin{table}[t!]
\begin{tabular*}{\columnwidth}{@{\extracolsep{\fill}}cccccc}
\toprule
 $r_s$ & $\rho_{\text{DM},\odot}$ & $\rho_{b,0}$ & $r_b$ & $\Sigma_{\odot}$ & $R_d$ \\
  $\left[\text{kpc}\right]$ & $[\text{GeV/cm}^3]$ & $[\text{GeV/cm}^3]$ &   $\left[\text{kpc}\right]$ & $[M_{\odot}/\text{pc}^2]$ &   $\left[\text{kpc}\right]$ \\
 \midrule
 30.36 & 0.19 & $1.83\times10^{4}$ & 0.092 & 57.9 & 3.2 \\
 \bottomrule
\end{tabular*}
\caption{The most likely values of the Galactic model parameters for the BCKM Galactic model obtained from a Markov Chain Monte Carlo
analysis from galactic rotation curve data\,\cite{Bhattacharjee:2012xm}. These parameters are largely insensitive to $z_d$, which is fixed to 340\,\text{pc}\,\cite{1998ApJ...492..495F}.}
\label{tab:gal-model}
\end{table}

%%%%%%%%%%%%%%%%%%%%%%%%%%%%%%%%%%%%%%%%%
\section{Velocity Distribution Functions}
%%%%%%%%%%%%%%%%%%%%%%%%%%%%%%%%%%%%%%%%%
\label{sec:VDF}

\subsection{Eddington formalism}
\label{subsec:edd-form}

\begin{figure}[]
\centering
\includegraphics[width=7.5 cm]{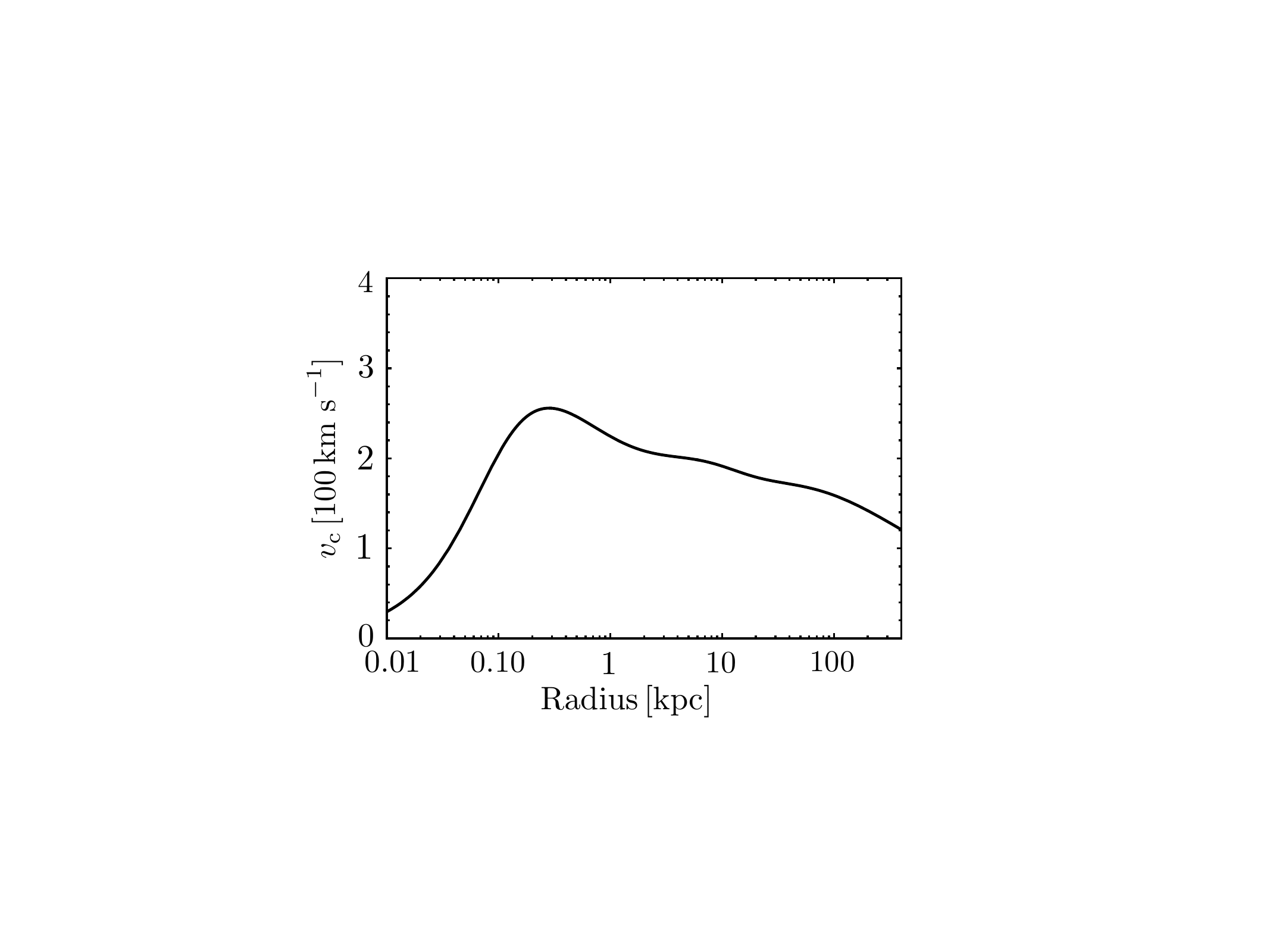}\;
\caption{
Circular speed $v_c(r)$ for the given Galactic model parameters listed in \cref{tab:gal-model}.}
\label{fig:circularspeed}
\end{figure}

\begin{figure}[]
\centering
\includegraphics[width=7.5 cm]{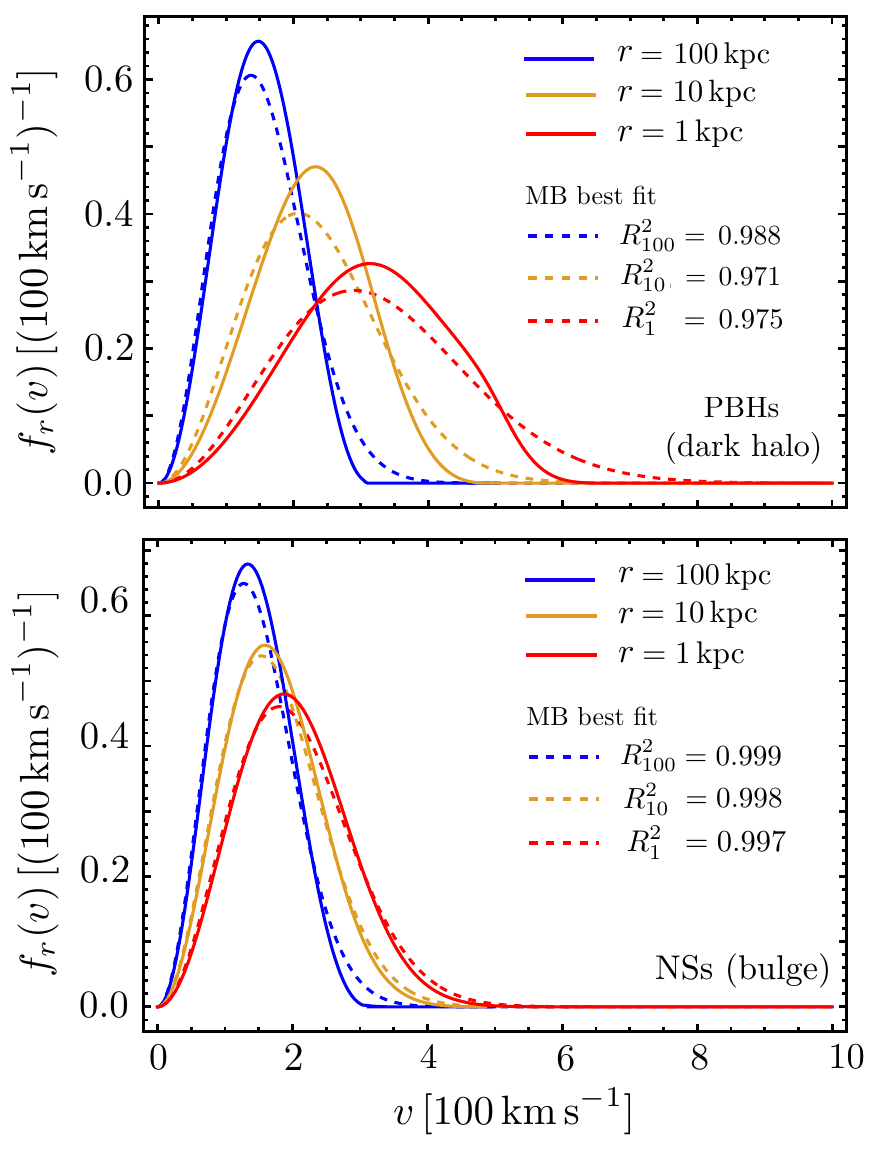}\;
\caption{Normalized speed distributions for PBHs in the dark halo (top) and NSs in the stellar bulge (bottom) in our galactic model at different Galactocentric radii (solid lines). Best-fits to Maxwell-Boltzmann 3-d isotropic velocity distributions are also shown (dashed lines).}
\label{fig:fvPBH}
\end{figure}

We use the Eddington formula to 
compute the velocity distribution functions (VDFs) of PBHs and NSs at a given position within the galactic dark halo and bulge, respectively, by linking their spatial density to the total galactic gravitational potential, $\Phi(r)$. 
In the spherical approximation\,\cite{Catena:2011kv}, this potential contains contributions from both dark and baryonic matter (BM) as
\begin{equation}
\begin{split}
\Phi(r) &= \Phi_{\text{DM}}(r) +\Phi_{\text{BM}}(r) \\ 
&\approx \int_0^r \frac{M_{\text{DM}}(r') + M_{\text{BM}}(r')}{r'^2/G_N}\, \mathrm{d}r'\,,
\end{split}
\end{equation}
where $M_{\text{DM}}$ and $M_{\text{BM}}$ are the total masses of the dark and baryonic components of the Galaxy, respectively. The total mass of the galaxy is then given by $M_\mathrm{total} = M_\mathrm{DM} + M_\mathrm{BM}$. For a spherically symmetric system composed of collisionless
components with isotropic VDFs, there exists a unique correspondence between the isotropic spatial density distribution $\rho(r)$ and its respective phase-space distribution function (PSDF) according to\,\cite{1916MNRAS..76..572E, 2008gady.book.....B}
\begin{equation}
\mathcal{F}(\xi) = \frac{1}{\sqrt{8}\pi^2} \left[ \int_0^{\xi} \frac{\mathrm{d}\Psi}{\sqrt{\xi-\Psi}}\frac{\mathrm{d}^2\rho}{\mathrm{d}\Psi^2} + \frac{1}{\sqrt{\xi}}\left(  \frac{\mathrm{d}\rho}{\mathrm{d}\Psi}\right)_{\Psi=0} \right]\,, 
\label{eq:PSDF}
\end{equation}
where $\xi = \Psi(r) - v^2 / 2$ is the relative energy defined in terms of the relative potential $\Psi(r) = -\Phi(r) + \Phi(r \xrightarrow{} \infty)$. From the PSDF in \cref{eq:PSDF}, the VDF at any radius $r$ is linked to the density profile via $\funop{f_r}(\vec{v}) = \mathcal{F}(\xi)/\rho_{\text{DM}}$. Each VDF satisfies the normalization condition 
\begin{equation}
    \int_0^\infty \funop{f_r}(v)\,\dl v = \int_0^\infty \int_\Omega \funop{f_r}(\vec{v}) v^2 \, \dl \Omega \, \dl v = 1\,,
\end{equation}
where $\funop{f_r}(v)$ is the normalized speed distribution (NSD).
The circular speed is calculated via ${v_{\text{c}}(r) = (G_N [M_{\text{DM}}(r)+M_{\text{BM}}(r)]/r)^{1/2}}$
for the Galactic model parameters of the Milky Way given in \cref{tab:gal-model}. We show this in \cref{fig:circularspeed}.
We calculate the VDF and NSD associated with PBHs in the Galactic dark halo by
setting the density profile to $\rho_{\text{DM}}$ in \cref{eq:rhodm}.
\cref{fig:fvPBH} (top) shows a selection of NSDs for PBHs in our galactic model at different radii $r$ (solid lines) as well as the best-fit Maxwell-Boltzmann (MB) 3-d isotropic velocity distributions (dashed lines) with the corresponding $R^2$ goodness-of-fit metric values quoted. Though best suited for linear models, this quantity provides us with a comparative metric with which to judge each of our MB fits, and we use it only for illustrative purposes.
As the radius decreases, the typical speeds move toward larger values and the distribution widens.  We have verified that our distributions agree with those computed by Ref.~\cite{Bhattacharjee:2012xm}. 
In the stellar bulge, we calculate the VDF and NSD associated with NSs by setting the density profile to $\rho_{b}(r)$ in \cref{eq:rhob}. \cref{fig:fvPBH} (bottom) shows a choice of NSDs for NSs in our galactic model at different radii $r$ (solid lines) as well as the best-fit MB 3-d isotropic velocity distributions (dashed lines) with the corresponding $R^2$ values. The NSDs are well-described by MB distributions at radius $r$ with a peak speed at $v_\text{c}(r)$.
In the disk, we cannot apply the Eddington formalism since the system is no longer spherically symmetric. In the limit of a razor-thin disk, we may assume that the VDF for NSs follows a bi-dimensional MB distribution in the $z=0$ galactic plane. We have numerically checked that the disk contribution to the final FRB rate is subdominant in comparison to that from the bulge, as previously pointed out in\,\cite{Kainulainen:2021rbg}. We therefore only focus on the Galactic bulge. 

%%%%%%%%%%%%%%%%%%%%%%%%%%%%%%%%%%%%%%%%%
\subsection{Galactic center}
%%%%%%%%%%%%%%%%%%%%%%%%%%%%%%%%%%%%%%%%%

Most galactic centers host supermassive black holes (SMBHs) with masses around $(10^6-10^8)\, M_{\odot}$\,\cite{Kormendy:2000cf}, which would have a strong impact on the dark and baryonic matter velocity distributions within certain characteristic galactic radii. This fact is relevant in the context of FRBs since the final FRB rate would receive a dominant contribution from  inner galactic regions, as mentioned in Ref.~\cite{Kainulainen:2021rbg}. Since it is computationally intensive to apply the Eddington formalism in regions very close to the Galatic center, we estimate this contribution by using the Jeans equation and reasonable assumptions. 
The so-called gravitational influence radius\,\cite{1972ApJ...178..371P} is defined as $r_h  \equiv G_N M_{\text{SMBH}}/\sigma^2$\,\cite{1972ApJ...178..371P}, where $\sigma$ is the stellar velocity dispersion of the host bulge. For the particular case of the Milky Way, we have $M_{\text{SMBH}}\approx 3.5\times 10^6\,M_{\odot}$ and $r_h$ extends until $\usim 3\,\mathrm{pc}$\,\cite{Safdi:2018oeu}. For galactic radii $r \gg r_h$, the SMBH presence can be safely neglected as we did in Sec. \ref{subsec:edd-form}. For $r \lesssim r_h$, we may estimate the velocity distributions for NSs and PBHs by solving the Jeans equation. We take both number densities to follow power-laws with an index $\bar{\gamma}$ and assume circular velocities $v_c(r) = (G_N M_{\text{total}}(r)/r)^{-1/2}$, where $M_{\text{total}}(r)$ is the total Galactic mass within a radius $r$. We obtain\,\cite{Dehnen:2006cm, Safdi:2018oeu}
\begin{equation}
\sigma_{i} \sim  123\,\mathrm{km\,s^{-1}}\left(1+\bar{\gamma}\right)^{-1/2}\left( \frac{M_{\text{total}}(r^*)}{r^*}\right)^{1/2}\,,
\end{equation}
where $r^*=1\,\text{pc}$, $\sigma_{i}$ is the 3-d velocity dispersion of an isotropic MB velocity distribution,  $i \in (\text{NS},\,\text{PBH})$, and we have taken
$\bar{\gamma} = 1$. The relative velocity dispersion is readily calculated, for example, as $\sigma_{\text{NS-PBH}} = \sqrt{\sigma_{\mathrm{NS}}^2 + \sigma_{\text{PBH}}^2}$.  

%%%%%%%%%%%%%%%%%%%%%%%%%%%%%%%%%%%%%%%%%
\section{Relative velocity distribution functions}
%%%%%%%%%%%%%%%%%%%%%%%%%%%%%%%%%%%%%%%%%
\label{Mcarlo}

The FRB rate is highly sensitive to the distribution of relative speeds between PBHs and NSs. While the NS-NS relative speed distribution is analytically tractable since their speeds follow an MB profile to a good approximation, those for NS-PBH and PBH-PBH must be found numerically. This is because the PBH VDF, $f_\mathrm{PBH}(\vec{v})$, is itself found numerically via \cref{eq:PSDF}. To retrieve these distributions, we proceed via a series of Monte Carlo (MC) simulations.
We begin by sampling $N_\mathrm{MC} = 10^6$ velocities from the relevant velocity distributions, $f_\mathrm{PBH/NS}(\vec{v})$. To find the distribution of relative speeds, we compute the magnitudes of the differences of these vectors to arrive at the relevant distribution function, $\funop{f_\mathrm{NS-PBH}}(v_\mathrm{rel})$ or $\funop{f_\mathrm{PBH-PBH}}(v_\mathrm{rel})$. We then employ Gaussian kernel density estimators to provide us with continuous functions of $f$ in terms of $v_\mathrm{rel}$.
\begin{figure}[]
\centering
\includegraphics[width=7.7 cm]{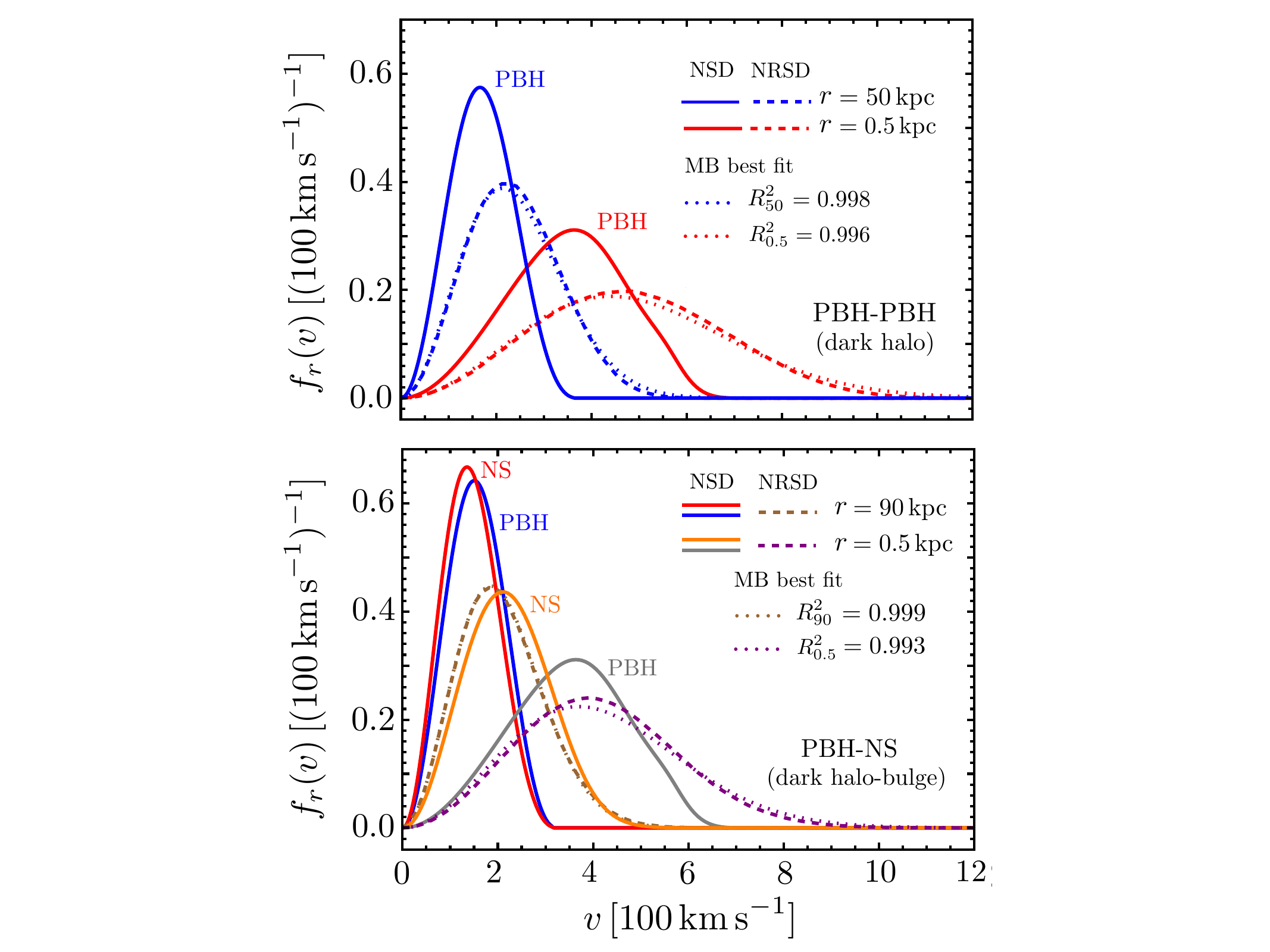}\;
\caption{Normalized relative speed distributions (NRSDs) between PBHs in the dark halo (top) and between PBHs and NSs in the stellar bulge (bottom) in our galactic model at different Galactocentric radii (dashed). In addition, the corresponding  best-fits to Maxwell-Boltzmann velocity distribution functions (dotted) are shown. The normalized speed distributions (NSDs) for PBHs (dark halo) and NSs (bulge) are also shown (solid).}
\label{fig:fvPBHNS}
\end{figure}
\begin{figure}[]
\centering
\includegraphics[width=8.5 cm]{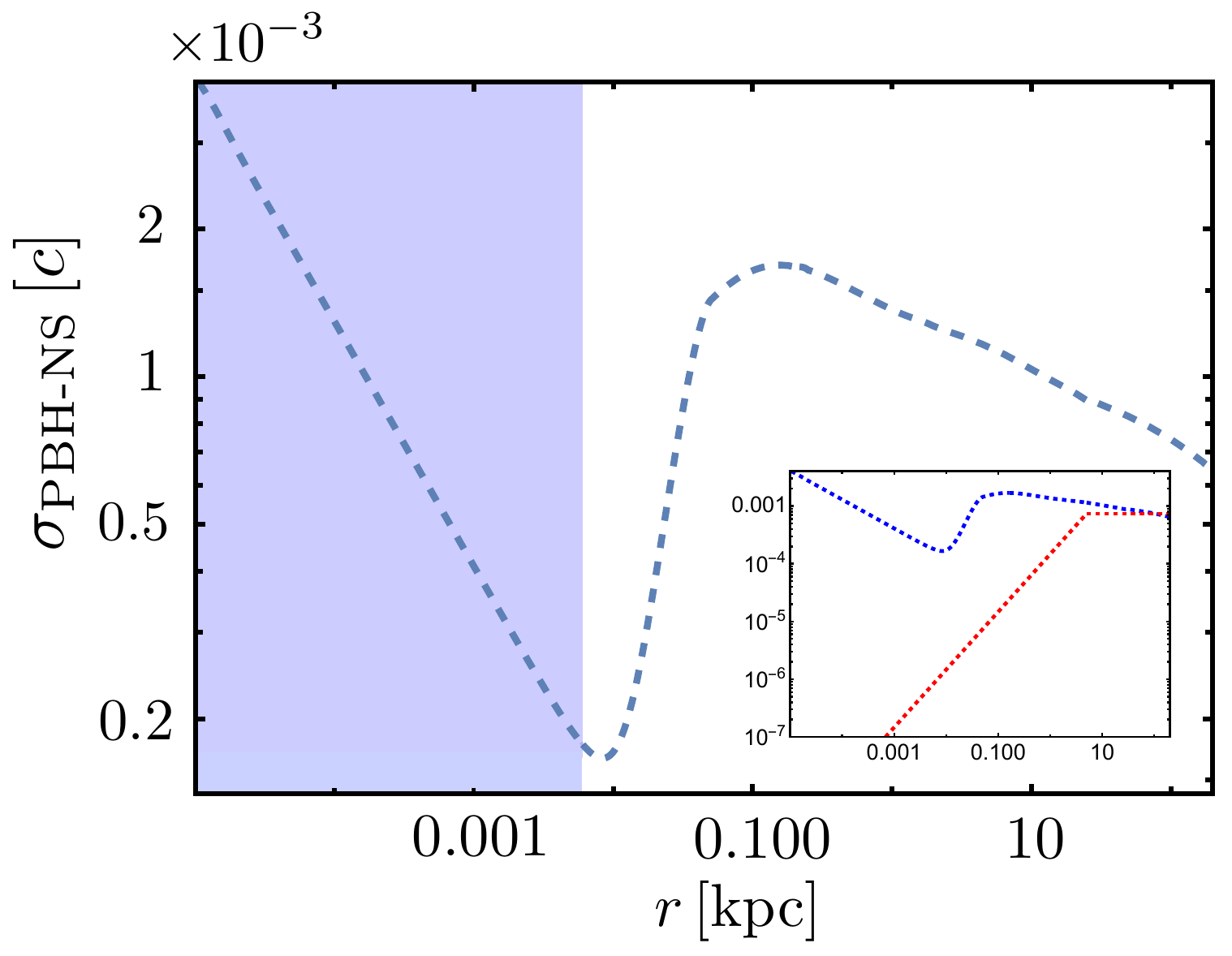}\;
\caption{Three-dimensional relative velocity dispersion  in light velocity units through the galaxy between PBHs and NSs in the bulge, $\sigma_{\text{PBH-NS}}$. The shaded zone indicates the inner galactic region, where we used the Jeans equation. The outer regions use the Eddington formalism, where we have fitted our $f_{\text{PBH-NS}}(v_{\text{rel}})$ to the closest MB 3-d isotropic velocity distributions. In the lower right corner, the red dashed curve shows the $\sigma_{\text{PBH-NS}}$ used in Refs.\,\cite{Abramowicz_2018,   Kainulainen:2021rbg}.}
\label{fig:galacticsigma}
\end{figure}
Figure \ref{fig:fvPBHNS} shows the normalized relative speed distributions 
$f_{\text{PBH-PBH}}$ (top) and $f_{\text{PBH-NS}}$ (bottom) for a given choice of galactic radii. In both cases, while solid lines refer to $f_r(v)$, the dashed lines show the corresponding $f_r(v_{\text{rel}})$. We observe that the normalized relative speed distributions tend to widen and peak at a higher speed in comparison to the original distributions. We also fit the relative speed distributions to MB profiles as in Sec.\,\ref{sec:gal-model}, finding excellent agreement despite the speed distributions of individual species, especially those of PBHs, departing from MB profiles. 
Neutron stars in the bulge have normalized speed distributions close to an MB profile, as Fig.\,\ref{fig:fvPBH} evinces. This explains why
the corresponding normalized relative speed distributions also closely follow MB distributions. 
Figure \ref{fig:galacticsigma} shows the three-dimensional relative velocity dispersion in light velocity units through the galaxy between PBHs and NSs in the bulge, $\sigma_{\text{PBH-NS}}$, as obtained in our fits to MB profiles. 
As explained in Sec.\,\ref{sec:VDF}, while we use the Jeans equation for regions close to the Galactic center under the presence of the central SMBH (blue shaded region), we use the Eddington formalism for the outer ones. 
The presence of the SMBH has a large impact on $\sigma_{\text{PBH-NS}}$, leading to 
$\sigma_{\text{PBH-NS}} \sim r^{-1/2}$ for $r \lesssim r_h$. This behavior departs from that assumed in Refs.\,\cite{Abramowicz_2018,   Kainulainen:2021rbg}, where the 3-D relative velocity dispersion was taken to be proportional to $r$ in the galactic inner regions. This has a significant impact on the predicted FRB rate since the PBH capture from NSs is highly sensitive to $\sigma_{\text{PBH-NS}}$. We will return to this point in Sec.\,\ref{Sec:RD}.
\vspace{0.4cm}

%%%%%%%%%%%%%%%%%%%%%%%%%%%%%%%%%%%%%%%%%
\section{Fast Radio Bursts Rate Calculation}
%%%%%%%%%%%%%%%%%%%%%%%%%%%%%%%%%%%%%%%%%
\label{Sec.FRBCal}

 After a PBH is captured by an NS, the PBH swallows the NS, leading to the formation of a black hole (BH) and the burst emission of the initial NS magnetic field energy. Such a burst is the source of FRBs. To compute this rate, we follow a similar approach to that in Ref.\,\cite{Kainulainen:2021rbg}. However, we introduce crucial improvements with respect to the galactic model and the corresponding velocity distributions of PBHs and NSs. 
We take a PBH all-DM scenario with a nonmonochromatic mass function with
a mean PBH mass $\overline{M}_{\text{PBH}}$.
The DM halo is composed of PBHs with $10^{-14}M_{\odot} \leq \overline{M}_{\text{PBH}} \leq 10^{-11}M_{\odot}$. As NSs ($M_N = 1.5 M_{\odot}$) are produced from the core-collapse of massive stars in the galactic disk and bulge, they will collide with PBHs, leading to a population of BHs ($M_0 = 1.5 M_{\odot}$). For consistency, we also consider a population of stellar black holes
($M_1 = 10 M_{\odot}$) coming from stellar evolution. Table \ref{tab:two} summarizes the species involved in our scenario\,\footnote{\textcolor{black}{We have used an extended lognormal mass function as $f(M_\mathrm{PBH}) = 1/(\sqrt{2\pi}\sigma)\exp[-(\text{ln}(M_{\text{PBH}}/M_c))^2/(2\sigma^2)]$, where $f(M_{\text{PBH}}) \equiv \frac{M_{\text{PBH}}}{\Omega_{\text{DM}}}\frac{\mathrm{d}\Omega_{\text{PBH}}}{\mathrm{d}M_{\text{PBH}}}$ is the differential mass function of the PBH DM fraction and $\overline{M}_{\text{PBH}}=M_c \exp(\sigma^2/2)$ for a spectrum centered at $M_c$ with a standard deviation equal to $\sigma$. For example, for $M_c \simeq 6\times10^{-14}\, M_{\odot}$, $\sigma \simeq 1.9$, we have $\overline{M}_{\text{PBH}} =  4\times10^{-13}\,M_{\odot}$.
We have tested that the actual width of the distribution has a small effect on the final results even taking $ f(M_{\text{PBH}})\delta(\overline{M}_{\text{PBH}}-M_c)$ as the PBH mass function}. The fact that the PBH masses in the PBH-all DM window are much smaller than the NS mass makes it such that the actual PBH mass does not have any effect on the gravitational cross section, \cref{eq:Cij}.}. 
As initial conditions, we set all
radial number densities to zero, but $n_P(t=0,\,r)=\rho_{\text{DM}}(r)/M_P$. We proceed to evolve in time all $n_{i}(t,\,r)$ 
by solving the following system of first-order  differential equations
\,\cite{Kainulainen:2021rbg}: 
\begin{align}
\small
&\dot{n}_N(t, r)  = - F_{\text{cap/coll}} C^{NP}(r)n_N(t,r)n_{P}(t,r) \nonumber \\
&-\sfrac{1}{2}C^{NN}(r)n_{N}(t, r)n_{N}(t, r)- C^{N0}(r)n_N(t, r) n_0(t, r)\,
\nonumber \\
&- C^{N1}(r)n_N(t, r) n_1(t, r)+ K^N(r)\,,
\label{NS} \\[0.5ex]
&\dot{n}_{P}(t,r) = - F_{\text{cap/coll}} C^{NP}(r)n_N(t,r) n_{P}(t,r) 
\nonumber \\ 
&-\sfrac{1}{2}C^{PP}(r)n_{P}(t,r)n_{P}(t,r) - C^{P0}(r)n_{P}(t,r)n_0(t,r)\,
\nonumber\\ 
&- C^{P1}(r)n_{P}(t,r)n_1(t,r)\,,
\label{p1} \\[0.5ex]
&\dot{n}_{0}(t,r) =  F_{\text{cap/coll}} C^{NP}(r)n_N(t,r) n_{P}(t,r)\,
\nonumber \\ 
&-C^{01}(r)n_0(t,r) n_{1}(t,r)-\sfrac{1}{2} C^{00}(r)n_0(t,r) n_{0}(t,r)\,,
\label{n0} \\[0.5ex]
&\dot{n}_{1}(t,r) = -\sfrac{1}{2}C^{11}(r)n_1(t,r)n_1(t,r) + K^1(r)\,,
\label{n1}
\end{align}
where $K^N(r)$ and $K^1(r)$ are the total creation rate for NSs and stellar BHs, respectively, $F_{\text{cap/coll}}$ is the ratio between the capture and collision rates for NS-PBH encounters, and $C^{ij}$ is the effective collision cross section averaged over the relative velocity between the $i$ and $j$ species. Collisions between the N species would likely form black holes, which can add to the 0 species population---that is, we may add the term $(1/2)C^{NN}n_N n_N$ on the right side of \cref{n0}. On the other hand, collisions between the P species may be approximately included in the monotonic-mass population $n_P$. This is akin to eliminating the term $(1/2)C^{PP}n_Pn_P$ in \cref{p1}. We have numerically checked that these changes do not appreciably impact our results.\footnote{ The rate of PP collisions should be larger than what is estimated in this work when considering PBH binary formation in the early Universe \cite{Sasaki:2018dmp}. However, under the approximation that the PP merger in the late Universe would form a PBH with a mass of the order of the original PBH species, our results would still hold to a reasonable approximation.} We model the creation rates $K^i(r)$, with $i \in \{N,\,1\}$, as $K^i(r) \equiv \bar{K}^i (\rho_b(r) / M_\mathrm{total})$, where $\bar{K}^1 = \bar{K}^N \in [0.007,\,0.063]\,\mathrm{yr^{-1}}$ are the creation rates over the entire galaxy. The ratio $F_{\text{cap/coll}}$ is calculated in\,\cite{Genolini:2020ejw} for comparing the initial PBH energy with all sources of energy dissipation during an NS-PBH encounter. Crucially, depending on the 
relative velocity between both astrophysical objects ($\sigma_{\text{PBH-NS}}$), such a ratio takes the form
\begin{align}
F_{\text{cap/coll}} &\approx \textcolor{black}{ 10^{-5}} \times
\left( \frac{{\color{black} \SI{300}{\kilo\meter\per\second}}}{\sigma_{\text{PBH-NS}}} \right)^2 \left( \frac{\textcolor{black}{M_{P}}}{\textcolor{black}{10^{-12}\,M_{\odot}}} \right) \mathcal{C}[X]\,,\label{fap/coll}
\end{align}
where
\begin{equation}
\mathcal{C}[X]\approx\left\lbrace\begin{array}{c c c}
&1\,\,\,&\text{for}\,\,\, X \lesssim 10\,,
\\ 
&X^{-1}\,\,\,&\text{for}\,\,\, 10\lesssim X \lesssim 1000\,,
\\
&X^{-5/7}\,\,\,&\text{for}\,\,\, X \gtrsim 10000\,,
\end{array}\right.
\end{equation}
with 
\begin{equation}
X \equiv 2\times 10^{-4}\left( \frac{10^{-12}M_{\odot}}{M_P} \right)^{-1} \left(\frac{\SI{300}{\kilo\meter\per\second}}{\sigma_{\text{PBH-NS}}}  \right)^2\,.
\end{equation}
\begin{table}[t!]
\begin{center}
\begin{tabular*}{\columnwidth}{llll}
\toprule
(Index) - Type & of astrophysical & species in the & galaxy \\
\midrule
(N) NS & $M_{N}=1.5 M_{\odot}$ &  $ n_{N}(t,\,r)$ & \hspace{-0.45cm} $R_N=10\,\text{km}$ \\
(P) PBH & $M_{P} = \overline{M}_{\text{PBH}}$ &  $n_{P}(t,\,r)$ & \hspace{-0.3cm}$R_{P} = R_s(M_{P})$ \\
(0) Light BH & $M_0=1.5 M_{\odot}$ & $ n_{0}(t,\,r)$ & \hspace{-0.25cm}$R_0=R_s(M_0)$  \\
(1) Stellar BH & $M_1 =10 M_{\odot}$ &  $ n_{1}(t,\,r)$ & \hspace{-0.25cm}$R_1=R_s(M_1)$\\
\bottomrule
\end{tabular*}
\caption{All astrophysical species of interest in the galaxy. Here, 
$M_i$, $n_i$, $R_i$, and $R_s(M_i)$ denote the mass, number density,  radius and Schwarzschild radius (if appropriate) of the $i$th species. The mean PBH mass in the all-DM scenario is denoted by $\overline{M}_{\text{PBH}}$.}
\label{tab:two}
\end{center}
\end{table}
 The effective collision cross section averaged over the relative velocities between $i$ and $j$ species reads as
 \begin{align}
C^{ij} &= \pi (R_i + R_j)^2 \nonumber \\
&\times\int\left(1 + \frac{2 G_N (M_i + M_j)}{(R_i + R_j)v^2_{\text{rel}}}  \right) v_{\text{rel}}f(\vec{v}_{\text{rel}})\,\dl^3 \vec{v}_{\text{rel}} \,,
\label{eq:Cij}
 \end{align}
where $f(\vec{v}_{\text{rel}})$ will depend on the interacting species. 
The system of equations (\ref{NS})-(\ref{n1}) is solved in  the galactic bulge
using the RVDFs obtained as indicated in Sec.\,\ref{sec:VDF}, where we assume that light and stellar BHs behave as NSs insofar as their velocity distributions are concerned. Since $\sigma_{\text{PBH-NS}}$ from Eq.\,(\ref{fap/coll}) were assumed in Ref.\,\cite{Genolini:2020ejw} to follow an MB 3-d isotropic velocity distribution, we fit our $f_{\text{PBH-NS}}(v_{\text{rel}})$ and $f_{\text{PBH-PBH}}(v_{\text{rel}})$  obtained numerically in the outer regions of the Galaxy to the closest MB distributions for consistency.

%%%%%%%%%%%%%%%%%%%%%%%%%%%%%%%%%%%%%%%%%
\section{Enhanced DM density at Galactic centers}
\label{sec:enhanced-density}
%%%%%%%%%%%%%%%%%%%%%%%%%%%%%%%%%%%%%%%%%

\textcolor{black}{Astrophysical observations indicate that almost all large galaxies host supermassive black holes at their centers\,\cite{Kormendy:1995er, Kormendy:2013dxa}}. Apart from the effect on the NS and PBHs velocity distributions, we expect that the formation and presence of central SMBHs would dynamically affect both baryonic and DM densities in the most inner galactic regions.  Since the FRB rate is dominated by inner galactic regions, a significant  overdensity associated with NSs and/or PBHs at galactic centers may boost the FRB rate\,\cite{Kainulainen:2021rbg}.
Here, we estimate the FRB rate arising from the sphere of gravitational influence around central SMBHs in spiral galaxies. This rate should be the order of the whole galactic rate. Although we use the Milky Way as a benchmark model for spiral galaxies, we strive to be as general as possible.
At galactic centers, NSs could hold a steeply-rising stellar density at radii $r \lesssim r_h$, with $r_h \approx 3\,\text{pc}$\,\cite{Safdi:2018oeu} in the case of the Milky Way. If the typical relaxation time is short enough, via a two-body  interaction, stars would exchange orbital energy leading to the formation of a Bahcall-Wolf density cusp\,\cite{1976ApJ...209..214B} around the SMBH. Numerical simulations\,\cite{Freitag:2006qf} shows that there should be around $10^4$ NSs within 1 pc around $\text{Sagittarius A}^*$ after $\usim 13\,\text{Gyr}$ holding a number density that may be approximated by an ``eta-model"\,\cite{1993MNRAS.265..250D, 1994AJ....107..634T, Safdi:2018oeu} with $\gamma = 1.3$, e.g.
\begin{equation}
n_{N}(r) = \frac{\eta A_{\text{NS}}}{4\pi r_b^3}\left(\frac{r}{r_b} \right)^{\eta-3}\left(1+\frac{r}{r_b} \right)^{-(\eta+1)}\,,
\label{eq:eta}
\end{equation}
where $\eta = 3-\gamma$, $r_b \approx 28\, \text{pc}/(2\eta-1)$,
and $A_{\text{NS}} = 749076$. 
Similarly, DM may form a collisional cusp near galactic centers via the Bahcall-Wolf mechanism. Even though DM is considered to be collisionless, they would be scattered off by stars\,\cite{2004JETP...98....1I, PhysRevLett.92.201304, PhysRevLett.93.061302} leading to a number density close to the SMBH as $n_{P}(r) \sim r^{-\gamma}$, with $\gamma=3/2$. This feature would be valid even in the case that the galactic center undergoes a previous scouring by a binary SMBH\,\cite{Merritt:2006mt}. 
On the other hand, if the SMBH is considered to grow adiabatically at the galactic center of the DM halo holding an initial singular power-law number density cusp, $n_{P}\sim r^{-\gamma}$ with $0<\gamma<2$ as suggested by numerical simulations\,\cite{Navarro:1995iw, 1998ApJ...499L...5M}, DM may form spikes at galactic centers being a steep function of the galactic radius as $n_{P}(r) \sim r^{-\gamma_{\text{sp}}}$ with $2.25\leq\gamma_{\text{sp}}\leq 2.5$\,\cite{Gondolo:1999ef, Bertone:2002je}. Even though the presence of such spikes requires particular initial conditions and  they can be later weakened or destroyed\,\cite{Ullio:2001fb}, we will still consider them for completeness to estimate the effect on the FRB rate. 
The actual form for the cuspy density profile at galactic centers, $n_{P,\,\text{cusp}}(r)$, that we consider takes the form
\begin{equation}
n_{P,\,\text{cusp}}(r)=\rho_{\text{DM}}(0.1r_h)M_P^{-1}\left(\frac{0.1r_h}{r}\right)^{3/2}\,,\\
\end{equation}
where $\rho_{\text{DM}}(0.1r_h)$ is obtained from Eq.\,(\ref{eq:rhodm}),
%$\rho_{\text{DM}}(0.1r_h) = \rho_{\text{DM},\odot} (R_0/0.1r_h)$, 
and the valid radii range is $4 R_s(M_{\text{SBH}}) \leq r \leq 0.1\,r_h$. 
For the spiky density profile, $n_{p,\,\text{sp}}(r)$, we follow\,\cite{Gondolo:1999ef} and take the form
\begin{equation}
n_{P,\,\text{sp}}(r) = \frac{\rho_{\text{DM,}\odot}}{M_P}\left(\frac{R_0}{R_{\text{sp}}}\right)^{\gamma}\left(1-\frac{4R_s(M_{\text{SBH}})}{r} \right)^3\left(  \frac{R_{\text{sp}}}{r}\right)^{\gamma_{\text{sp}}}\,,  
\end{equation}
where $4 R_s(M_{\text{SBH}}) \leq r \leq R_{\text{sp}}$ and $\gamma_{\text{sp}}=(9-2\gamma)/(4-\gamma)$ with
$R_{\text{sp}}= \alpha_{\gamma}R_0(M_{\text{SBH}}/\rho_{\text{DM},\odot}R_0^3)^{1/(3-\gamma)}$ and
${0\leq \gamma \leq 2}$, respectively. Here $\alpha_{\gamma} \sim (10^{-3}-10^{-1})$ depends on the $\gamma$-value and is 
obtained numerically. 
To estimate the velocity dispersion for NSs and PBHs at galactic centers, we solve the Jeans equation \,\cite{Dehnen:2006cm} as before and use the fact that both number densities are taken to follow power-laws with certain indices $\bar{\gamma}$ and assume circular velocities $v_c(r) \sim r^{-1/2}$ to obtain
\begin{equation}
\sigma_{i} \sim 123\,\mathrm{km\,s^{-1}}\left(1+\bar{\gamma}\right)^{-1/2}\left( \frac{1\,\text{pc}}{r}\right)^{1/2} \hfill \text{for}\,r \lesssim 3\,\text{pc}\,,
\end{equation}
where $\sigma_{i}$ is the 3-d velocity dispersion of a isotropic MB velocity distribution,  $i \in (\text{NS},\text{PBH})$,
$\bar{\gamma} = (1.3,3/2,\gamma_{\text{sp}})$ as appropriate, and  $M_{\text{total}}\approx M_{\text{SBH}}$ within the radius of interest. The relative velocity dispersion is readily calculated as $\sigma_{\text{PBH-NS}} = \sqrt{\sigma_\mathrm{NS}^2 + \sigma_{\text{PBH}}^2}$ as
well as the effective collision cross section average over $\sigma_{\text{PBH-NS}}$, (\cref{eq:Cij}), as\,\cite{Nurmi:2021xds}
\begin{equation}
C^{\mathrm{NP}} \approx \sqrt{8\pi R_N^4\sigma^2_{\text{PBH-NS}}/3}  \left( 1+\frac{3G_N M_N}{R_{N}\sigma^2_{\text{PBH-NS}}}\right)\,.
\end{equation}
We estimate the FRB rate associated with a density enhanced environment at galactic centers by integrating the inner region as
\begin{equation}
\text{FRB}^{\text{center}}_{\text{sp,\,cusp}} = \int^{r_i}_{4R_s}4\pi r^2 F_{\text{cap/coll}}(r)C^{NP}(r) n_N(r)n_{P,i}(r)\,\dl r\,, 
\label{eq:frb-inner}
\end{equation}
where the NS number density is given by Eq.\,(\ref{eq:eta}) and the PBH number density is chosen to hold a cuspy or spiky profile, e.g. $n_{P,\,i} = (n_{P,\,\text{cusp}},\, n_{P,\,\text{sp}})$ and ${r_i=(0.1r_h,\, \text{min}(r_h,\, R_{\text{sp}}))}$, respectively.

%%%%%%%%%%%%%%%%%%%%%%%%%%%%%%%%%%%%%%%%%%%%%%%
\section{Results and Discussion}
\label{Sec:RD}
%%%%%%%%%%%%%%%%%%%%%%%%%%%%%%%%%%%%%%%%%%%%%%%
\subsection{Main results}

\begin{figure}[]
\centering
\includegraphics[width=8.7 cm]{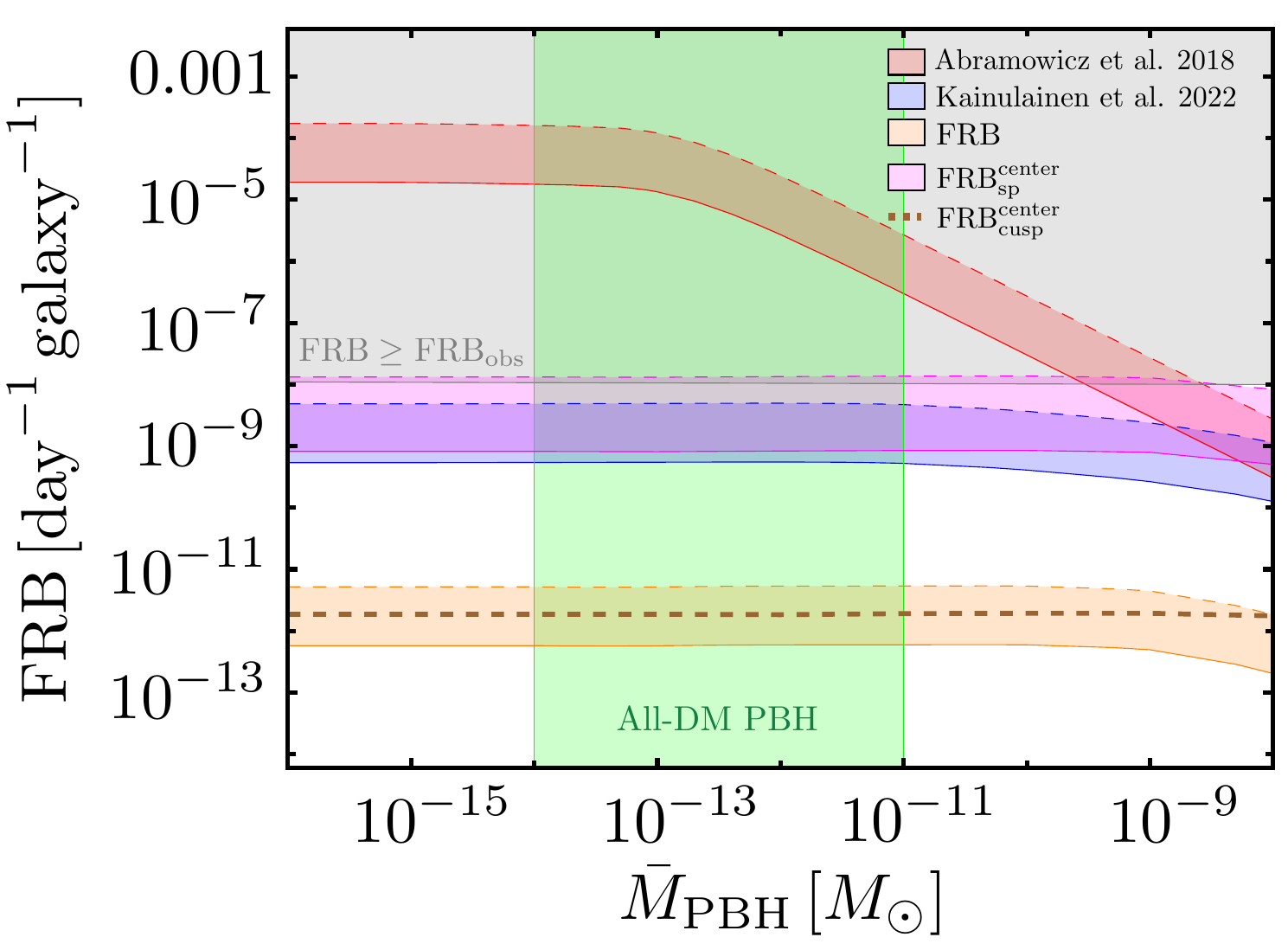}\;
\caption{Predicted FRB rate in a typical spiral galaxy for $0.007\,\text{yr}^{-1} \leq \bar{K}^N = \bar{K}^1 \leq 0.063\, \text{yr}^{-1}$
 obtained in Abramowicz \textit{et al.} 2018\,\cite{Abramowicz_2018}, Kainulainen \textit{et al.} 2022\,\cite{Kainulainen:2021rbg}, and this work (\textcolor{black}{red}, blue and orange shaded regions, respectively). We also estimate the FRB rate from galactic centers considering a cuspy density profile for NSs ($n_{N,\,\text{cusp}}\sim r^{-1.3}$) together to a (1) cuspy ($n_{P,\,\text{cusp}}\sim r^{-3/2}$) or (2) spiky ($n_{P,\,\text{sp}}\sim r^{-(2.26,2.33)}$) PBH density profile (brown dashed line and purple shaded region). The green shaded region highlights the PBH mass window where PBHs can constitute all of DM. The gray shaded region indicates an FRB rate equal to or greater than the actual observed FRB rate, $\mathrm{FRB}_\mathrm{obs} \sim 10^{-8}\,\text{day}^{-1}\,\text{galaxy}^{-1}$\,\cite{Champion:2015pmj, Petroff:2019tty}.}
\label{fig:FRBresult}
\end{figure}

Numerical results based on previous sections are summarized in Fig.\,\ref{fig:FRBresult}. The current number of FRBs per day and per galaxy in a typical spiral galaxy in terms of the PBH mass is shown as an orange shaded region, where $\bar{K}^N = \bar{K}^1 = 0.063\,\text{yr}^{-1}$ (dashed
orange line) and $0.007\,\text{yr}^{-1}$
(solid orange line). This rate is calculated using a similar procedure to that in Ref.~\cite{Kainulainen:2021rbg}, but taking the Milky Way as the galactic model and calculating the
relative velocity distributions between different astrophysical species (NSs, PBHs, and BHs) more accurately. For the outer galactic regions, we use the Eddington inversion formula and a series of MC simulations and MB-profile fits, while for the inner galactic regions we use the Jeans equation. In the parameter space of interest, where all DM can be composed of PBHs, $10^{-14}M_{\odot}\lesssim M_{P} \lesssim 10^{-11}M_{\odot}$ (green shaded region), the predicted FRB rate is about 3-4 orders of magnitude smaller than the observed FRB rate, $ \mathrm{FRB}_\mathrm{obs} \sim 10^{-8}\,\text{day}^{-1}\,\text{galaxy}^{-1}$\,\cite{Champion:2015pmj, Petroff:2019tty}. 
This result improves that from Ref.\,\cite{Kainulainen:2021rbg} (blue shaded region), where  the calculation of relative velocities between PBH and NS in the galactic inner region is oversimplified. In detail, the assumed model was the piece-wise function $\sigma_{\text{PBH-NS}}=(\SI{200}{\kilo\meter\per\second})\, r/r_b$ for $r\leq r_b$ and $\sigma_{\text{PBH-NS}}=(\SI{200}{\kilo\meter\per\second})$ for $r> r_b$, where $r_b = 5\,\text{kpc}$. As the red dashed line in the lower right corner of Fig.\,\ref{fig:galacticsigma} shows, the relative velocity dispersion is highly underestimated in such a case, leading to an overestimation of the FRB rate, where the PBH capture from NSs strongly depends on $\sigma_{\text{PBH-NS}}$ via Eq.\,(\ref{fap/coll}). Indeed, the presence of an SMBH at galactic centers fixes the total mass of the galaxy in the most inner regions, increasing the PBH-NS relative velocity dispersion, $\sigma_{\text{PBH-NS}}\sim r^{-1/2}$, and suppressing the PBH capture, $F_{\text{cap/coll}}\sim \sigma_{\text{PBH-NS}}^{-2}$, as one approaches the center. 
In addition, as was first pointed out in Ref.~\cite{Kainulainen:2021rbg}, we have shown that the very first estimates for the FRB rate due to the capture of PBHs performed in Ref.\,\cite{Abramowicz_2018} (red shaded region) largely overestimated the FRB signal due to NS--PBH captures. This is because they assumed all collisions lead to a PBH capture ($F_\mathrm{cap/coll} = 1$), arriving at the controversial conclusion that the PBH all-DM scenario is ruled out. 
Besides the effect on the NS-PBH velocity distributions, it is expected that the formation and presence of central SMBHs in spiral galaxies would dynamically affect their densities in the most inner region. To obtain an estimate of such an effect, we have calculated the rate for FRBs arising from the sphere of gravitational influence due to a central SMBH.
Considering cuspy NS and PBH density profiles, i.e. $n_N(r) \sim r^{-1.3}$ and $n_{P,\,\text{cusp}}\sim r^{-3/2}$, respectively, does not impact the final FRB rate for an order-of-magnitude estimate (dashed brown line). However, considering a cuspy NS density profile but a spiky PBH density profile, i.e. $n_{P,\,\text{cusp}}\sim r^{-\gamma_{\text{sp}}}$ with $\gamma_{\text{sp}}=(9-2\gamma)/(4-\gamma)$ and $0 \leq \gamma \leq 2$, the FRB rate may increase by several orders of magnitude. The shaded purple region corresponds to $2.26 \lesssim \gamma_{\text{sp}} \lesssim 2.33$. We see that the predicted signal would be consistent with the observed rate for an order-of-magnitude estimate for a spiky enough DM profile.  
Nevertheless, this last statement should be taken with caution. Even though galactic centers may hold a DM spike, several conditions need to be satisfied for this to be true. Only if the SMBH is placed just at the center of the DM halo, a tiny initial mass may grow adiabatically in the preexistent DM halo to its final mass at the present day. If the formation of the SMBH occurs away from the center and undergoes an inspiral motion, the central DM spike is weakened as $n_P(r) \sim r^{-1/2}$ or even transformed into a depletion rather than an enhancement\,\cite{Ullio:2001fb}. 

\textcolor{black}{We conclude that the PBH all-DM hypothesis is not in tension with 
the observed rate of FRB and, indeed, may be consistent with it given an order-of-magnitude estimate in the case that most galaxies host sufficiently spiky PBH central densities due to the presence of a central SMBH}.

\vspace{-0.2 cm}
\subsection{Future directions}

\textcolor{black}{We comment on the original assumption made in Ref.\,\cite{Abramowicz_2018}, which we have followed, regarding the specific use of spiral galaxies to estimate the FRB rate due to NS-PBH mergers. From this discussion, we will address the main future directions to be taken regarding the present work.}
\textcolor{black}{We have assumed that all galaxies at $z \lesssim 1$, where most FRBs are located\,\cite{Champion:2015pmj, Petroff:2019tty, Piratova:2024}, are spirals for the purposes of computing the FRB signal and that the Milky Way is an average spiral. \textcolor{black}{On the latter, the Milky Way is considered to be a structurally typical spiral galaxy\,\cite{Goodwin:1997ys, Cirelli:2024ssz}. Indeed, rotation characteristics of spiral galaxies and mass distributions are similar among the various types of spiral galaxies, as was shown in  Ref.\,\cite{2017PASJ...69R...1S}. Even in the case of sizeable mass variations, we do not expect a significant change in our main conclusions. We will return to this point at the end of this section}.  On the former, we note that the FRB rate mostly arises from the galactic inner region. It can thus be approximately written as
\begin{equation}
    \mathrm{FRB}_{\text{S}} \sim  N_{\mathrm{N},\text{S}}  \left[ F_{\text{cap/coll},\text{S}}\, C^\mathrm{NP}_{\text{S}} n_{\mathrm{P},\text{S}}\right]_{\text{center}}\,,
    \label{eq:FRBrateprop}
\end{equation}
where quantities inside the squared parenthesis are averaged over the central region,  the subscript ``S" emphasizes the fact that we are focusing on a typical spiral galaxy, and \textcolor{black}{$N_N$ refers to the total number of neutron stars}. To be more general, we may 
estimate the total FRB rate per day and per galaxy, $\text{FRB}_{\text{S+}}$, by extending Eq.\,(\ref{eq:FRBrateprop}) to all
different types of galaxies as follows:
\begin{equation}
\begin{split}
\text{FRB}_{\text{S+}}& \sim \frac{1}{N} \sum_{i} N_{i}\text{FRB}_{\text{i}}\,,\\
& \sim \frac{1}{N}\left( 1 + \sum_{i \neq \textcolor{black}{S}} \alpha_i \frac{\text{FRB}_{i}}{\text{FRB}_{\text{S}}}  \right) N_S \, \text{FRB}_{\text{S}}\,,
\end{split}
\label{eq:truerate}
\end{equation}
where the subindex $i$ runs over elliptical (E), irregular (I), and dwarf (D) galaxies \textcolor{black}{in the second line of Eq.\,(\ref{eq:truerate})},  $N \equiv N_{\text{S}}+N_{\text{E}}+N_{\text{I}}+N_{\text{D}}$ with $N_i$ the number of the $i$-type of galaxy for $z \lesssim 1$, and $\alpha_i \equiv N_i/N_S$ being the fraction of $i$-type galaxies with respect to spirals. The rate in Eq.\,(\ref{eq:truerate}), therefore, depends on the collision cross section and the fraction of captured PBHs per PBH-NS collision, which both themselves depend on the relative velocity dispersion $\sigma_\mathrm{PBH-NS}$, the number of neutron stars in a galaxy, and the number density of PBHs. To estimate the contribution of other morphologies on the FRB rate, we must thus comment on (i) the number of galaxies present at $z \lesssim 1$, (ii) the $F_{\text{cap/coll}}\, C^\mathrm{NP}$ factor, (iii) the number density of NS, and (iv) the DM density profile for all galaxy types. We consider spiral, elliptical, irregular, and dwarf galaxies in a series of order-of-magnitude estimates.}

\textcolor{black}{
On the first point, dwarf galaxies constitute the largest population of galaxies in the Universe\,\cite{Crnoj:2021,2022ARA&A..60...73S, 2022Annibali}. For larger and brighter galaxies, spirals make up $\usim 60\%$ of the makeup of the Universe, followed by ellipticals ($\usim 10\%$) and then irregulars ($\usim 4\%$)\,\cite{Loveday:1996xn}. Focusing then on dwarfs, the known population of Milky Way satellite galaxies have grown with time due to the increase of galactic surveys. With the Dark Energy Survey\,\cite{2015ApJ...807...50B, 2015ApJ...813..109D}, other DECam surveys such as SMASH5\,\cite{2015ApJ...804L...5M}) and DELVE7\,\cite{2021ApJ...920L..44C}, ATLAS\,\cite{2016MNRAS.463..712T}, and Gaia\,\cite{2019MNRAS.488.2743T}, the number of confirmed candidate satellites has risen to more than $60$\,\cite{Crnoj:2021}.  A similar population of faint dwarfs has been observed to surround M31 (see, e.g.,\,\cite{2009ApJ...705..758M}). Even though most of the current reduced knowledge about dwarf galaxies is based on local observations, successful wide-field searches and spectroscopic surveys around the Milky Way are currently pushing the boundaries of our knowledge beyond the Local Group\,\cite{2016ApJ...823...19C, 2017ApJ...850..109B}. For example, the SAGA survey, which focuses on constructing a statistical sample of satellite systems around Milky Way-like galaxies, reported between $1$ and $9$ satellites per host galaxy\,\cite{2021ApJ...907...85M}. Thus, we roughly expect there to be $\usim 10^2$ more dwarfs than spirals and $\usim 10^{-1}$ as many ellipticals and irregulars.}

\textcolor{black}{On the second point, a key result of this work is that the velocity dispersion nontrivially depends on the assumed dark matter and stellar density profiles, and the predicted FRB rate is highly sensitive to it, requiring numerical computations to properly estimate. The  effective collision cross section average over $\sigma_{\text{PBH-NS}}$ is dominated by the gravitational enhancement so that $C^{\text{NP}} \sim \sigma_{\text{PBH-NS}}^{-1}$. On the other hand, we have $F_{\text{cap/coll}} \sim \sigma_{\text{PBH-NS}}^{-2}$ for $X \approx  1$ in Eq.\,(\ref{fap/coll}). A change in the dark and stellar galactic profiles will affect $\sigma_{\text{PBH-NS}}$ and, as a result, the $F_{\text{cap/coll}}C^{\text{NP}}$. However, presently, we will take this factor to be of the same order for all types of galaxies. We have numerically tested the velocity dispersion of dark and baryonic matter using the BKCM model, but doubling the DM halo NFW concentration.  The $\sigma_{\text{PBH-NS}}$ value changes by a few percent}. 

\textcolor{black}{
On the third point, we can estimate the number of NSs in each galaxy type by using the predicted rate of core-collapse supernovae, which are predominantly responsible for their production\,\cite{2021MNRAS.500.1071M}. Integrating the morphology-specific rates over the age of the Universe, we retrieve the number of NSs. We find that spirals and ellipticals should contain roughly the same order of NS, $\usim 10^8\,\mathrm{NS}$, with irregulars containing an order of magnitude fewer,  $\usim 10^{7}\,\mathrm{NS}$. For dwarfs, it is unclear whether\,\cite{2021MNRAS.500.1071M} considered them; however, we can argue for their NS number from the perspective of relative galactic masses. Dwarf galaxies typically have masses $10^7-10^9 M_{\odot}$, which are two orders of magnitude smaller than typical spiral galaxy masses ($10^9-10^{12} M_{\odot}$). Assuming that the populations of stars are roughly equivalent in both dwarfs and spirals, which may be a lower estimate since dwarfs usually possess an older population of stars \cite{Andrade:2023jul}, we can then approximate the number of NS in dwarfs to be $\usim 10^{-2}$ that of spirals.}

\textcolor{black}{Finally, on the fourth point, we comment on the DM density in each morphology. In cold dark matter (CDM), the structure of virialized dark matter halos shows a spherically averaged density profile (the NFW profile) according to  $N$-body simulations\,\cite{1996ApJ...462..563N}. We used such a profile in Eq.\,(\ref{eq:rhodm}), Sec.\,\ref{sec:gal-model}}.

\textcolor{black}{In elliptical galaxies, dark matter is an extremely common (probably ubiquitous) constituent of them\,\cite{Loewenstein:1999ue}.
For example, using x-ray observations (Chandra, XMM, and Suzaku), the authors in Ref.\,\cite{Buote:2011zk} show that an NFW profile for DM halos is able to broadly fit available data within the observational errors for $7$ elliptical galaxies in the mass range of $10^{12}\,M_{\odot}-10^{13}\,M_{\odot}$, with predicted virial concentrations slightly larger than 10, consistent with observations. Such concentrations are close to that of the Milky Way BCKM-model used in Sec.\,\ref{sec:gal-model} considering the upper range at the 68$\%$ of confidence level (see Table 1 in Ref.\,\cite{Bhattacharjee:2012xm}).}

\textcolor{black}{Regarding dwarf galaxies, we currently know that they constitute the largest population of galaxies in the Universe\,\cite{Crnoj:2021}, but they are also very faint and difficult to observe\,\cite{2013MNRAS.429.3068T}. 
On the one hand, they typically have lower luminosity, mass, and size than typical spiral galaxies, but they have proportionally more dark matter. They are considered to be dark matter-dominated structures, but the presence of a cuspy DM central profile remains unclear due to possible simulation shortcomings\,\cite{2002A&A...385..816D, 2013MNRAS.429.3068T} and/or systematic uncertainties\,\cite{vandenBosch:1999ka, 2001MNRAS.325.1017V}. Authors in Ref.\,\cite{Simon:2003tf} analyze DM density profiles of nearby dwarf galaxies, reducing systematic uncertainties\,\cite{Simon:2003xu}. Their results indicate that observational evidence does not support the thesis that dwarf galaxies hold a universal DM profile. While NGC 2976 %(pure disk system with $\sim10^{9}\,M_{\odot}$) 
 shows a nearly constant-density DM core with $\rho_{\text{DM}}\sim 0.1 M_{\odot}/\text{pc}^3$ at 1 pc, NGC 4605 holds a density profile with an intermediate feature between a core and cuspy shape ($\rho_{\text{DM}} \propto r^{-0.65}$), and NGC 5963 
 a DM density closely consistent with an NFW profile with a concentration of 20.}

\textcolor{black}{Irregular galaxies show unusual shapes, which makes astronomers believe that they could come from galaxy interactions or collisions. Thus, it is plausible to consider that they could share the characteristics and compositions of the original galaxies, such as star types and dark matter content.}

\textcolor{black}{
From the considerations above, to a first approximation, we take the amount of DM available for NS capture in elliptical, spiral and irregular galaxies to be all of the same order. For the peculiar case of dwarf galaxies, we take a conservative approach and consider them to have on average 1 or 2  orders of magnitude more DM.}

\textcolor{black}{
Now, we are ready to estimate Eq.\,(\ref{eq:truerate}).
Taking into consideration the relative population of galaxies with respect to spiral galaxies, the number of NSs, and the DM densities, we have $\alpha_{\text{E}}=\alpha_{\text{I}} \sim 10^{-1}$, $\alpha_{\text{D}}\sim 10^2$, $\text{FRB}_{\text{E}}\sim \text{FRB}_{\text{S}}$, 
$\text{FRB}_{\text{I}}=10^{-1}\text{FRB}_{\text{S}}$, and $\text{FRB}_{\text{D}}\sim (10^{-1}-10^0)\,\text{FRB}_{\text{S}}$.  We compare Eq.\,(\ref{eq:truerate}) with the FRB rate signal predicted by us shown as an orange band in Fig.\,\ref{fig:FRBresult}, obtaining
$\text{FRB}_{\text{S+}}/\text{FRB}_\text{S} \sim 10^{-1}-10^0$. This estimate tells us,  to a first approximation, that considering spiral galaxies as a representative galactic candidate for FRB signal is a reasonable idea.} 

\textcolor{black}{As we previously commented on, the Milky Way is a good representative spiral galaxy in terms of properties such as its mass distribution, dark halo, and rotation curves\,\cite{2017PASJ...69R...1S}.
With a mass of approximately} $10^{12}\,M_{\odot}$, \textcolor{black}{the Milky Way is among the heavier spiral galaxies. However, to explore the impact that more massive spirals could have on our prediction, we may consider a small fraction of "super heavy" spiral galaxies, which contribute with an FRB rate per day and per galaxy equal to $\mathrm{FRB}_{\text{SS}}$.  Modeling all galaxies as spirals, the modified FRB signal,  $\mathrm{FRB}_{\text{mod}}$, would take the form}
\begin{equation}
    \mathrm{FRB}_{\text{mod}} \sim  (1-\epsilon)\mathrm{FRB}_{\text{S}} + \epsilon\mathrm{FRB}_{\text{SS}}\,,
    \label{eq:FRBSandbeyond}
\end{equation}
\textcolor{black}{where $\epsilon$ is the fraction of super spiral galaxies  with respect to all spiral galaxies. The super spiral galaxies will dominate the signal if $\epsilon > \text{FRB}_{\text{S}}/(\text{FRB}_{\text{S}} + \text{FRB}_{\text{SS}})$. For instance, if the typical FRB rate from super spiral galaxies is around $10^2$ times larger than that from typical spiral galaxies, they will dominate the signal for $\epsilon \gtrsim 0.01$.}

\textcolor{black}{To get a sense of how much our prediction based on the Milky Way-like signal would change, let us consider the conservative case for which super spiral galaxies constitute $1\%$ of all spiral galaxies and are typically one order of magnitude heavier in mass than typical spirals. Using the BCKM-model\,\cite{Bhattacharjee:2012xm}, we know that, for the Milky Way, the fraction of DM and NSs with respect to the total galactic mass at the virial radius are about $0.92$ and $1.5 \times 10^{-3}$, respectively, where we have considered a total of $\usim 10^9$ NSs\,\cite{Kainulainen:2021rbg} and $M_N = 1.5\,M_{\odot}$. Populating our heavy super spiral galaxies in the same proportion and taking the extreme case in which their volume is of the order as that of the Milky Way,  
the new averaged number densities  in them are 
$\bar{n}^{\text{SS}}_{N} \sim 10 \bar{n}_{N}$ and
$\bar{n}^{\text{SS}}_{P} \sim 10 \bar{n}_{P}$, where
$\bar{n}_{N}$ and $\bar{n}_{P}$
are the typical averaged number densities for the Milky Way. 
From Eq.\,(\ref{eq:FRBSandbeyond}), we may see that the modified FRB rate will increase by an $\mathcal{O}(1)$ factor in comparison with that from typical spiral galaxies. More dramatically, if we suppose that super spiral galaxies are much older than the Milky Way and contain around 10 times more NSs than this if they would have a Milky Way mass, 
the factor for the total FRB rate would instead be of $\mathcal{O}(10)$. Our rough estimates, which do not account for more complicated variables such as changes in typical NS-PBH velocities, are sufficient to approximate the total predicted modified FRB rate.} 

As a result, 
since the observed FRB rate is larger than our prediction by 2 or 3 orders of magnitude, we may properly conclude that the tension between such a rate and the PBH all-DM hypothesis does not exist.

\textcolor{black}{Finally, we comment on the purple shaded region on Fig.\,\ref{fig:FRBresult}, where we conclude that sufficiently spiky PBH central densities would make the predicted FRB rate consistent with the observed one. Since dwarf galaxies are much lighter than spiral galaxies, the presence of a central SMBH, which would enhance the sharpness of DM profiles, is not guaranteed. However, there has been observational evidence of interest. For example, the most distant dwarf spheroidal galaxy from the Milky Way, Leo I ($ \usim 8 \times 10^7\,M_{\odot}$), would host an SMBH ( $M_{\text{SMBH}} \sim 10^6\,M_{\odot}$) according to dynamical evidence\,\cite{Pacucci:2023eaf}. Such a black hole mass is around 3 orders of magnitude larger than the expected mass. If dwarf galaxies hold enough spiky DM central densities, their effect on the total FRB rate may be comparable with those from ``spiky"-spiral galaxies. Future work is needed to settle down this compelling scenario}.    

%%%%%%%%%%%%%%%%%%%%%%%%%%%%%%%%%%%%%%%%%%%%%
\section{Acknowledgments}
%%%%%%%%%%%%%%%%%%%%%%%%%%%%%%%%%%%%%%%%%%%%%

E.D.S. and D.W.P.A. thank Andrew J.~Long (Rice University) for his comments and suggestions. E.D.S. and D.W.P.A. thank Megan Reiter (Rice University) for discussions on the velocity distribution functions in galaxies. E.D.S. thanks Kimmo Kainulainen (University of Jyv$\ddot{\text{a}}$skl$\ddot{\text{a}}$) and Sami Nurmi (University of Jyv$\ddot{\text{a}}$skl$\ddot{\text{a}}$ and University of Helsinki) for useful comments about this manuscript. \textcolor{black}{D.A. thanks Jennifer Mead for helpful discussions toward the end of this project.} E.D.S. was supported in part by the National Science Foundation under Award No.~PHY-2114024 \textcolor{black}{during his stay in Rice University, U.S.A., as postdoctoral associate and Rice Academy Junior Fellow (2022-2023)}. \textcolor{black}{E.D.S. thanks the Physics Department at
the Metropolitan University of Educational Sciences, Chile, where part of this work was done}.
D.W.P.A. was supported by the National Science Foundation under award 2209444. 
\\
\noindent
$^*$\href{mailto:enrico.d.schiappacasse@rice.edu}{Enrico.Schiappacasse@uss.cl}\\
$^\dagger$\href{mailto:dorian.amaral@rice.edu}{dorian.amaral@rice.edu}\\
    
\bibliography{FRBPBH} 

\begin{thebibliography}{92}%
\makeatletter
\providecommand \@ifxundefined [1]{%
 \@ifx{#1\undefined}
}%
\providecommand \@ifnum [1]{%
 \ifnum #1\expandafter \@firstoftwo
 \else \expandafter \@secondoftwo
 \fi
}%
\providecommand \@ifx [1]{%
 \ifx #1\expandafter \@firstoftwo
 \else \expandafter \@secondoftwo
 \fi
}%
\providecommand \natexlab [1]{#1}%
\providecommand \enquote  [1]{``#1''}%
\providecommand \bibnamefont  [1]{#1}%
\providecommand \bibfnamefont [1]{#1}%
\providecommand \citenamefont [1]{#1}%
\providecommand \href@noop [0]{\@secondoftwo}%
\providecommand \href [0]{\begingroup \@sanitize@url \@href}%
\providecommand \@href[1]{\@@startlink{#1}\@@href}%
\providecommand \@@href[1]{\endgroup#1\@@endlink}%
\providecommand \@sanitize@url [0]{\catcode `\\12\catcode `\$12\catcode
  `\&12\catcode `\#12\catcode `\^12\catcode `\_12\catcode `\%12\relax}%
\providecommand \@@startlink[1]{}%
\providecommand \@@endlink[0]{}%
\providecommand \url  [0]{\begingroup\@sanitize@url \@url }%
\providecommand \@url [1]{\endgroup\@href {#1}{\urlprefix }}%
\providecommand \urlprefix  [0]{URL }%
\providecommand \Eprint [0]{\href }%
\providecommand \doibase [0]{http://dx.doi.org/}%
\providecommand \selectlanguage [0]{\@gobble}%
\providecommand \bibinfo  [0]{\@secondoftwo}%
\providecommand \bibfield  [0]{\@secondoftwo}%
\providecommand \translation [1]{[#1]}%
\providecommand \BibitemOpen [0]{}%
\providecommand \bibitemStop [0]{}%
\providecommand \bibitemNoStop [0]{.\EOS\space}%
\providecommand \EOS [0]{\spacefactor3000\relax}%
\providecommand \BibitemShut  [1]{\csname bibitem#1\endcsname}%
\let\auto@bib@innerbib\@empty
%</preamble>
\bibitem [{\citenamefont {Peebles}(2015)}]{Peebles:2013hla}%
  \BibitemOpen
  \bibfield  {author} {\bibinfo {author} {\bibfnamefont {P.~J.~E.}\
  \bibnamefont {Peebles}},\ }\bibfield  {title} {\enquote {\bibinfo {title}
  {{Dark Matter}},}\ }\href {\doibase 10.1073/pnas.1308786111} {\bibfield
  {journal} {\bibinfo  {journal} {Proc. Natl. Acad. Sci. U.S.A.}\ }\textbf
  {\bibinfo {volume} {112}},\ \bibinfo {pages} {2246} (\bibinfo {year}
  {2015})}\BibitemShut {NoStop}%
\bibitem [{\citenamefont {Freese}(2017)}]{Freese:2017idy}%
  \BibitemOpen
  \bibfield  {author} {\bibinfo {author} {\bibfnamefont {K.}~\bibnamefont
  {Freese}},\ }\bibfield  {title} {\enquote {\bibinfo {title} {{Status of Dark
  Matter in the Universe}},}\ }\href {\doibase 10.1142/S0218271817300129}
  {\bibfield  {journal} {\bibinfo  {journal} {Int. J. Mod. Phys.}\ }\textbf
  {\bibinfo {volume} {1}},\ \bibinfo {pages} {325} (\bibinfo {year}
  {2017})}\BibitemShut {NoStop}%
\bibitem [{\citenamefont {Bertone}\ and\ \citenamefont
  {Hooper}(2018)}]{Bertone:2016nfn}%
  \BibitemOpen
  \bibfield  {author} {\bibinfo {author} {\bibfnamefont {G.}~\bibnamefont
  {Bertone}}\ and\ \bibinfo {author} {\bibfnamefont {D.}~\bibnamefont
  {Hooper}},\ }\bibfield  {title} {\enquote {\bibinfo {title} {{History of dark
  matter}},}\ }\href {\doibase 10.1103/RevModPhys.90.045002} {\bibfield
  {journal} {\bibinfo  {journal} {Rev. Mod. Phys.}\ }\textbf {\bibinfo {volume}
  {90}},\ \bibinfo {pages} {045002} (\bibinfo {year} {2018})}\BibitemShut
  {NoStop}%
\bibitem [{\citenamefont {Jungman}\ \emph {et~al.}(1996)\citenamefont
  {Jungman}, \citenamefont {Kamionkowski},\ and\ \citenamefont
  {Griest}}]{Jungman:1995df}%
  \BibitemOpen
  \bibfield  {author} {\bibinfo {author} {\bibfnamefont {Gerard}\ \bibnamefont
  {Jungman}}, \bibinfo {author} {\bibfnamefont {Marc}\ \bibnamefont
  {Kamionkowski}}, \ and\ \bibinfo {author} {\bibfnamefont {Kim}\ \bibnamefont
  {Griest}},\ }\bibfield  {title} {\enquote {\bibinfo {title} {{Supersymmetric
  dark matter}},}\ }\href {\doibase 10.1016/0370-1573(95)00058-5} {\bibfield
  {journal} {\bibinfo  {journal} {Phys. Rept.}\ }\textbf {\bibinfo {volume}
  {267}},\ \bibinfo {pages} {195--373} (\bibinfo {year} {1996})}\BibitemShut
  {NoStop}%
\bibitem [{\citenamefont {Preskill}\ \emph {et~al.}(1983)\citenamefont
  {Preskill}, \citenamefont {Wise},\ and\ \citenamefont
  {Wilczek}}]{PRESKILL1983127}%
  \BibitemOpen
  \bibfield  {author} {\bibinfo {author} {\bibfnamefont {J.}~\bibnamefont
  {Preskill}}, \bibinfo {author} {\bibfnamefont {M.~B.}\ \bibnamefont {Wise}},
  \ and\ \bibinfo {author} {\bibfnamefont {F.}~\bibnamefont {Wilczek}},\
  }\bibfield  {title} {\enquote {\bibinfo {title} {Cosmology of the invisible
  axion},}\ }\href {\doibase https://doi.org/10.1016/0370-2693(83)90637-8}
  {\bibfield  {journal} {\bibinfo  {journal} {Phys. Lett. B}\ }\textbf
  {\bibinfo {volume} {120}},\ \bibinfo {pages} {127} (\bibinfo {year}
  {1983})}\BibitemShut {NoStop}%
\bibitem [{\citenamefont {Abbott}\ and\ \citenamefont
  {Sikivie}(1983)}]{ABBOTT1983133}%
  \BibitemOpen
  \bibfield  {author} {\bibinfo {author} {\bibfnamefont {L.F.}\ \bibnamefont
  {Abbott}}\ and\ \bibinfo {author} {\bibfnamefont {P.}~\bibnamefont
  {Sikivie}},\ }\bibfield  {title} {\enquote {\bibinfo {title} {A cosmological
  bound on the invisible axion},}\ }\href {\doibase
  https://doi.org/10.1016/0370-2693(83)90638-X} {\bibfield  {journal} {\bibinfo
   {journal} {Phys. Lett. B}\ }\textbf {\bibinfo {volume} {120}},\ \bibinfo
  {pages} {133} (\bibinfo {year} {1983})}\BibitemShut {NoStop}%
\bibitem [{\citenamefont {Dine}\ and\ \citenamefont
  {Fischler}(1983)}]{DINE1983137}%
  \BibitemOpen
  \bibfield  {author} {\bibinfo {author} {\bibfnamefont {M.}~\bibnamefont
  {Dine}}\ and\ \bibinfo {author} {\bibfnamefont {W.}~\bibnamefont
  {Fischler}},\ }\bibfield  {title} {\enquote {\bibinfo {title} {The
  not-so-harmless axion},}\ }\href {\doibase
  https://doi.org/10.1016/0370-2693(83)90639-1} {\bibfield  {journal} {\bibinfo
   {journal} {Phys. Lett. B}\ }\textbf {\bibinfo {volume} {120}},\ \bibinfo
  {pages} {137} (\bibinfo {year} {1983})}\BibitemShut {NoStop}%
\bibitem [{\citenamefont {Kim}\ and\ \citenamefont
  {Carosi}(2010)}]{Kim:2008hd}%
  \BibitemOpen
  \bibfield  {author} {\bibinfo {author} {\bibfnamefont {J.~E.}\ \bibnamefont
  {Kim}}\ and\ \bibinfo {author} {\bibfnamefont {G.}~\bibnamefont {Carosi}},\
  }\bibfield  {title} {\enquote {\bibinfo {title} {{Axions and the Strong CP
  Problem}},}\ }\href {\doibase 10.1103/RevModPhys.82.557} {\bibfield
  {journal} {\bibinfo  {journal} {Rev. Mod. Phys.}\ }\textbf {\bibinfo {volume}
  {82}},\ \bibinfo {pages} {557} (\bibinfo {year} {2010})},\ \bibinfo {note}
  {[Erratum: Rev.Mod.Phys. 91, 049902 (2019)]}\BibitemShut {NoStop}%
\bibitem [{\citenamefont {d'Enterria}(2021)}]{dEnterria:2021ljz}%
  \BibitemOpen
  \bibfield  {author} {\bibinfo {author} {\bibfnamefont {David}\ \bibnamefont
  {d'Enterria}},\ }\bibfield  {title} {\enquote {\bibinfo {title} {{Collider
  constraints on axion-like particles}},}\ }in\ \href@noop {} {\emph {\bibinfo
  {booktitle} {{Workshop on Feebly Interacting Particles}}}}\ (\bibinfo {year}
  {2021})\ \Eprint {http://arxiv.org/abs/2102.08971} {arXiv:2102.08971
  [hep-ex]} \BibitemShut {NoStop}%
\bibitem [{\citenamefont {Buchmueller}\ \emph {et~al.}(2017)\citenamefont
  {Buchmueller}, \citenamefont {Doglioni},\ and\ \citenamefont
  {Wang}}]{Buchmueller:2017qhf}%
  \BibitemOpen
  \bibfield  {author} {\bibinfo {author} {\bibfnamefont {Oliver}\ \bibnamefont
  {Buchmueller}}, \bibinfo {author} {\bibfnamefont {Caterina}\ \bibnamefont
  {Doglioni}}, \ and\ \bibinfo {author} {\bibfnamefont {Lian~Tao}\ \bibnamefont
  {Wang}},\ }\bibfield  {title} {\enquote {\bibinfo {title} {{Search for dark
  matter at colliders}},}\ }\href {\doibase 10.1038/nphys4054} {\bibfield
  {journal} {\bibinfo  {journal} {Nat. Phys.}\ }\textbf {\bibinfo {volume}
  {13}},\ \bibinfo {pages} {217} (\bibinfo {year} {2017})}\BibitemShut
  {NoStop}%
\bibitem [{\citenamefont {{Rybka (ADMX
  Collaboration)}}(2014)}]{2014PDU.....4...14R}%
  \BibitemOpen
  \bibfield  {author} {\bibinfo {author} {\bibfnamefont {Gray}\ \bibnamefont
  {{Rybka (ADMX Collaboration)}}},\ }\bibfield  {title} {\enquote {\bibinfo
  {title} {{Direct detection searches for axion dark matter}},}\ }\href
  {\doibase 10.1016/j.dark.2014.05.003} {\bibfield  {journal} {\bibinfo
  {journal} {Phys. Dark Universe}\ }\textbf {\bibinfo {volume} {4}},\ \bibinfo
  {pages} {14} (\bibinfo {year} {2014})}\BibitemShut {NoStop}%
\bibitem [{\citenamefont {Roszkowski}\ \emph {et~al.}(2018)\citenamefont
  {Roszkowski}, \citenamefont {Sessolo},\ and\ \citenamefont
  {Trojanowski}}]{Roszkowski:2017nbc}%
  \BibitemOpen
  \bibfield  {author} {\bibinfo {author} {\bibfnamefont {L.}~\bibnamefont
  {Roszkowski}}, \bibinfo {author} {\bibfnamefont {E.~M.}\ \bibnamefont
  {Sessolo}}, \ and\ \bibinfo {author} {\bibfnamefont {S.}~\bibnamefont
  {Trojanowski}},\ }\bibfield  {title} {\enquote {\bibinfo {title} {{WIMP dark
  matter candidates and searches\textemdash{}current status and future
  prospects}},}\ }\href {\doibase 10.1088/1361-6633/aab913} {\bibfield
  {journal} {\bibinfo  {journal} {Rep. Prog. Phys.}\ }\textbf {\bibinfo
  {volume} {81}},\ \bibinfo {pages} {066201} (\bibinfo {year}
  {2018})}\BibitemShut {NoStop}%
\bibitem [{\citenamefont {Schumann}(2019)}]{Schumann:2019eaa}%
  \BibitemOpen
  \bibfield  {author} {\bibinfo {author} {\bibfnamefont {Marc}\ \bibnamefont
  {Schumann}},\ }\bibfield  {title} {\enquote {\bibinfo {title} {{Direct
  Detection of WIMP Dark Matter: Concepts and Status}},}\ }\href {\doibase
  10.1088/1361-6471/ab2ea5} {\bibfield  {journal} {\bibinfo  {journal} {J.
  Phys. G}\ }\textbf {\bibinfo {volume} {46}},\ \bibinfo {pages} {103003}
  (\bibinfo {year} {2019})}\BibitemShut {NoStop}%
\bibitem [{\citenamefont {Hawking}(1971)}]{Hawking:1971ei}%
  \BibitemOpen
  \bibfield  {author} {\bibinfo {author} {\bibfnamefont {S.}~\bibnamefont
  {Hawking}},\ }\bibfield  {title} {\enquote {\bibinfo {title}
  {{Gravitationally collapsed objects of very low mass}},}\ }\href {\doibase
  10.1093/mnras/152.1.75} {\bibfield  {journal} {\bibinfo  {journal} {Mon. Not.
  R. Astron. Soc.}\ }\textbf {\bibinfo {volume} {152}},\ \bibinfo {pages} {75}
  (\bibinfo {year} {1971})}\BibitemShut {NoStop}%
\bibitem [{\citenamefont {{Chapline}}(1975)}]{1975Natur.253..251C}%
  \BibitemOpen
  \bibfield  {author} {\bibinfo {author} {\bibfnamefont {G.~F.}\ \bibnamefont
  {{Chapline}}},\ }\bibfield  {title} {\enquote {\bibinfo {title}
  {{Cosmological effects of primordial black holes}},}\ }\href {\doibase
  10.1038/253251a0} {\bibfield  {journal} {\bibinfo  {journal} {Nature
  (London)}\ }\textbf {\bibinfo {volume} {253}},\ \bibinfo {pages} {251}
  (\bibinfo {year} {1975})}\BibitemShut {NoStop}%
\bibitem [{\citenamefont {Inomata}\ \emph {et~al.}(2017)\citenamefont
  {Inomata}, \citenamefont {Kawasaki}, \citenamefont {Mukaida}, \citenamefont
  {Tada},\ and\ \citenamefont {Yanagida}}]{Inomata:2017okj}%
  \BibitemOpen
  \bibfield  {author} {\bibinfo {author} {\bibfnamefont {K.}~\bibnamefont
  {Inomata}}, \bibinfo {author} {\bibfnamefont {M.}~\bibnamefont {Kawasaki}},
  \bibinfo {author} {\bibfnamefont {K.}~\bibnamefont {Mukaida}}, \bibinfo
  {author} {\bibfnamefont {Y.}~\bibnamefont {Tada}}, \ and\ \bibinfo {author}
  {\bibfnamefont {T.~T.}\ \bibnamefont {Yanagida}},\ }\bibfield  {title}
  {\enquote {\bibinfo {title} {{Inflationary Primordial Black Holes as All Dark
  Matter}},}\ }\href {\doibase 10.1103/PhysRevD.96.043504} {\bibfield
  {journal} {\bibinfo  {journal} {Phys. Rev. D}\ }\textbf {\bibinfo {volume}
  {96}},\ \bibinfo {pages} {043504} (\bibinfo {year} {2017})}\BibitemShut
  {NoStop}%
\bibitem [{\citenamefont {Nakamura}(1998)}]{PhysRevLett.80.1138}%
  \BibitemOpen
  \bibfield  {author} {\bibinfo {author} {\bibfnamefont {T.~T.}\ \bibnamefont
  {Nakamura}},\ }\bibfield  {title} {\enquote {\bibinfo {title} {Gravitational
  lensing of gravitational waves from inspiraling binaries by a point mass
  lens},}\ }\href {\doibase 10.1103/PhysRevLett.80.1138} {\bibfield  {journal}
  {\bibinfo  {journal} {Phys. Rev. Lett.}\ }\textbf {\bibinfo {volume} {80}},\
  \bibinfo {pages} {1138} (\bibinfo {year} {1998})}\BibitemShut {NoStop}%
\bibitem [{\citenamefont {Takahashi}\ and\ \citenamefont
  {Nakamura}(2003)}]{Takahashi:2003ix}%
  \BibitemOpen
  \bibfield  {author} {\bibinfo {author} {\bibfnamefont {R.}~\bibnamefont
  {Takahashi}}\ and\ \bibinfo {author} {\bibfnamefont {T.}~\bibnamefont
  {Nakamura}},\ }\bibfield  {title} {\enquote {\bibinfo {title} {{Wave effects
  in gravitational lensing of gravitational waves from chirping binaries}},}\
  }\href {\doibase 10.1086/377430} {\bibfield  {journal} {\bibinfo  {journal}
  {Astrophys. J.}\ }\textbf {\bibinfo {volume} {595}},\ \bibinfo {pages} {1039}
  (\bibinfo {year} {2003})}\BibitemShut {NoStop}%
\bibitem [{\citenamefont {Carr~\textit{et al.}}(2021)}]{Carr:2020gox}%
  \BibitemOpen
  \bibfield  {author} {\bibinfo {author} {\bibfnamefont {Bernard}\ \bibnamefont
  {Carr~\textit{et al.}}},\ }\bibfield  {title} {\enquote {\bibinfo {title}
  {{Constraints on Primordial Black Holes}},}\ }\href {\doibase
  10.1088/1361-6633/ac1e31} {\bibfield  {journal} {\bibinfo  {journal} {Rep.
  Prog. Phys.}\ }\textbf {\bibinfo {volume} {84}},\ \bibinfo {pages} {116902}
  (\bibinfo {year} {2021})}\BibitemShut {NoStop}%
\bibitem [{\citenamefont {Inomata~et al.}(2018)}]{Inomata:2018cht}%
  \BibitemOpen
  \bibfield  {author} {\bibinfo {author} {\bibfnamefont {Keisuke}\ \bibnamefont
  {Inomata~et al.}},\ }\bibfield  {title} {\enquote {\bibinfo {title} {{Double
  inflation as a single origin of primordial black holes for all dark matter
  and LIGO observations}},}\ }\href {\doibase 10.1103/PhysRevD.97.043514}
  {\bibfield  {journal} {\bibinfo  {journal} {Phys. Rev. D}\ }\textbf {\bibinfo
  {volume} {97}},\ \bibinfo {pages} {043514} (\bibinfo {year}
  {2018})}\BibitemShut {NoStop}%
\bibitem [{\citenamefont {Fuller}\ \emph {et~al.}(2017)\citenamefont {Fuller},
  \citenamefont {Kusenko},\ and\ \citenamefont {Takhistov}}]{Fuller:2017uyd}%
  \BibitemOpen
  \bibfield  {author} {\bibinfo {author} {\bibfnamefont {G.~M.}\ \bibnamefont
  {Fuller}}, \bibinfo {author} {\bibfnamefont {A.}~\bibnamefont {Kusenko}}, \
  and\ \bibinfo {author} {\bibfnamefont {V.}~\bibnamefont {Takhistov}},\
  }\bibfield  {title} {\enquote {\bibinfo {title} {{Primordial Black Holes and
  $r$-Process Nucleosynthesis}},}\ }\href {\doibase
  10.1103/PhysRevLett.119.061101} {\bibfield  {journal} {\bibinfo  {journal}
  {Phys. Rev. Lett.}\ }\textbf {\bibinfo {volume} {119}},\ \bibinfo {pages}
  {061101} (\bibinfo {year} {2017})}\BibitemShut {NoStop}%
\bibitem [{\citenamefont {Katz}(2018)}]{Katz:2018xiu}%
  \BibitemOpen
  \bibfield  {author} {\bibinfo {author} {\bibfnamefont {J.~I.}\ \bibnamefont
  {Katz}},\ }\bibfield  {title} {\enquote {\bibinfo {title} {{Fast radio
  bursts}},}\ }\href {\doibase 10.1016/j.ppnp.2018.07.001} {\bibfield
  {journal} {\bibinfo  {journal} {Prog. Part. Nucl. Phys.}\ }\textbf {\bibinfo
  {volume} {103}},\ \bibinfo {pages} {1} (\bibinfo {year} {2018})}\BibitemShut
  {NoStop}%
\bibitem [{\citenamefont {Abramowicz}\ \emph {et~al.}(2018)\citenamefont
  {Abramowicz}, \citenamefont {Bejger},\ and\ \citenamefont
  {Wielgus}}]{Abramowicz_2018}%
  \BibitemOpen
  \bibfield  {author} {\bibinfo {author} {\bibfnamefont {M.~A.}\ \bibnamefont
  {Abramowicz}}, \bibinfo {author} {\bibfnamefont {M.}~\bibnamefont {Bejger}},
  \ and\ \bibinfo {author} {\bibfnamefont {M.}~\bibnamefont {Wielgus}},\
  }\bibfield  {title} {\enquote {\bibinfo {title} {Collisions of neutron stars
  with primordial black holes as fast radio bursts engines},}\ }\href {\doibase
  10.3847/1538-4357/aae64a} {\bibfield  {journal} {\bibinfo  {journal}
  {Astrophys. J.}\ }\textbf {\bibinfo {volume} {868}},\ \bibinfo {pages} {17}
  (\bibinfo {year} {2018})}\BibitemShut {NoStop}%
\bibitem [{\citenamefont {Totani}(2013)}]{Totani:2013lia}%
  \BibitemOpen
  \bibfield  {author} {\bibinfo {author} {\bibfnamefont {T.}~\bibnamefont
  {Totani}},\ }\bibfield  {title} {\enquote {\bibinfo {title} {{Cosmological
  Fast Radio Bursts from Binary Neutron Star Mergers}},}\ }\href {\doibase
  10.1093/pasj/65.5.L12} {\bibfield  {journal} {\bibinfo  {journal} {Publ.
  Astron. Soc. Jpn.}\ }\textbf {\bibinfo {volume} {65}},\ \bibinfo {pages}
  {L12} (\bibinfo {year} {2013})}\BibitemShut {NoStop}%
\bibitem [{\citenamefont {Lipunov}\ and\ \citenamefont
  {Pruzhinskaya}(2014)}]{Lipunov:2013axa}%
  \BibitemOpen
  \bibfield  {author} {\bibinfo {author} {\bibfnamefont {V.~M.}\ \bibnamefont
  {Lipunov}}\ and\ \bibinfo {author} {\bibfnamefont {M.~V.}\ \bibnamefont
  {Pruzhinskaya}},\ }\bibfield  {title} {\enquote {\bibinfo {title} {{Scenario
  Machine: fast radio bursts, short gamma-ray burst, dark energy and Laser
  Interferometer Gravitational-wave Observatory silence}},}\ }\href {\doibase
  10.1093/mnras/stu313} {\bibfield  {journal} {\bibinfo  {journal} {Mon. Not.
  R. Astron. Soc.}\ }\textbf {\bibinfo {volume} {440}},\ \bibinfo {pages}
  {1193} (\bibinfo {year} {2014})}\BibitemShut {NoStop}%
\bibitem [{\citenamefont {Yamasaki}\ \emph {et~al.}(2018)\citenamefont
  {Yamasaki}, \citenamefont {Totani},\ and\ \citenamefont
  {Kiuchi}}]{Yamasaki:2017hdr}%
  \BibitemOpen
  \bibfield  {author} {\bibinfo {author} {\bibfnamefont {S.}~\bibnamefont
  {Yamasaki}}, \bibinfo {author} {\bibfnamefont {T.}~\bibnamefont {Totani}}, \
  and\ \bibinfo {author} {\bibfnamefont {K.}~\bibnamefont {Kiuchi}},\
  }\bibfield  {title} {\enquote {\bibinfo {title} {{Repeating and Non-repeating
  Fast Radio Bursts from Binary Neutron Star Mergers}},}\ }\href {\doibase
  10.1093/pasj/psy029} {\bibfield  {journal} {\bibinfo  {journal} {Publ.
  Astron. Soc. Jpn.}\ }\textbf {\bibinfo {volume} {70}},\ \bibinfo {pages} {39}
  (\bibinfo {year} {2018})}\BibitemShut {NoStop}%
\bibitem [{\citenamefont {D.~Gupta}\ and\ \citenamefont
  {Saini}(2018)}]{DasGupta:2017uac}%
  \BibitemOpen
  \bibfield  {author} {\bibinfo {author} {\bibfnamefont {P.}~\bibnamefont
  {D.~Gupta}}\ and\ \bibinfo {author} {\bibfnamefont {N.}~\bibnamefont
  {Saini}},\ }\bibfield  {title} {\enquote {\bibinfo {title} {{Collapsing
  supra-massive magnetars: FRBs, the repeating FRB121102 and GRBs}},}\ }\href
  {\doibase 10.1007/s12036-017-9499-9} {\bibfield  {journal} {\bibinfo
  {journal} {J. Astrophys. Astron.}\ }\textbf {\bibinfo {volume} {39}},\
  \bibinfo {pages} {14} (\bibinfo {year} {2018})}\BibitemShut {NoStop}%
\bibitem [{\citenamefont {{Dai}}\ \emph {et~al.}(2017)\citenamefont {{Dai}},
  \citenamefont {{Wang}},\ and\ \citenamefont {{Yu}}}]{2017ApJ...838L...7D}%
  \BibitemOpen
  \bibfield  {author} {\bibinfo {author} {\bibfnamefont {Z.~G.}\ \bibnamefont
  {{Dai}}}, \bibinfo {author} {\bibfnamefont {J.~S.}\ \bibnamefont {{Wang}}}, \
  and\ \bibinfo {author} {\bibfnamefont {Y.~W.}\ \bibnamefont {{Yu}}},\
  }\bibfield  {title} {\enquote {\bibinfo {title} {{Radio Emission from Pulsar
  Wind Nebulae without Surrounding Supernova Ejecta: Application to FRB
  121102}},}\ }\href {\doibase 10.3847/2041-8213/aa6745} {\bibfield  {journal}
  {\bibinfo  {journal} {Astrophys. J. Lett.}\ }\textbf {\bibinfo {volume}
  {838}},\ \bibinfo {eid} {L7} (\bibinfo {year} {2017})}\BibitemShut {NoStop}%
\bibitem [{\citenamefont {Waxman}(2017)}]{Waxman:2017zme}%
  \BibitemOpen
  \bibfield  {author} {\bibinfo {author} {\bibfnamefont {E.}~\bibnamefont
  {Waxman}},\ }\bibfield  {title} {\enquote {\bibinfo {title} {{On the origin
  of fast radio bursts (FRBs)}},}\ }\href {\doibase 10.3847/1538-4357/aa713e}
  {\bibfield  {journal} {\bibinfo  {journal} {Astrophys. J.}\ }\textbf
  {\bibinfo {volume} {842}},\ \bibinfo {pages} {34} (\bibinfo {year}
  {2017})}\BibitemShut {NoStop}%
\bibitem [{\citenamefont {Zhang}(2018)}]{Zhang:2018csb}%
  \BibitemOpen
  \bibfield  {author} {\bibinfo {author} {\bibfnamefont {B.}~\bibnamefont
  {Zhang}},\ }\bibfield  {title} {\enquote {\bibinfo {title} {{Fast Radio Burst
  Energetics and Detectability from High Redshifts}},}\ }\href {\doibase
  10.3847/2041-8213/aae8e3} {\bibfield  {journal} {\bibinfo  {journal}
  {Astrophys. J. Lett.}\ }\textbf {\bibinfo {volume} {867}},\ \bibinfo {pages}
  {L21} (\bibinfo {year} {2018})}\BibitemShut {NoStop}%
\bibitem [{\citenamefont {Buckley}\ \emph {et~al.}(2021)\citenamefont
  {Buckley}, \citenamefont {Dev}, \citenamefont {Ferrer},\ and\ \citenamefont
  {Huang}}]{Buckley:2020fmh}%
  \BibitemOpen
  \bibfield  {author} {\bibinfo {author} {\bibfnamefont {J.~H.}\ \bibnamefont
  {Buckley}}, \bibinfo {author} {\bibfnamefont {P.~S.~B.}\ \bibnamefont {Dev}},
  \bibinfo {author} {\bibfnamefont {F.}~\bibnamefont {Ferrer}}, \ and\ \bibinfo
  {author} {\bibfnamefont {F.~P.}\ \bibnamefont {Huang}},\ }\bibfield  {title}
  {\enquote {\bibinfo {title} {{Fast radio bursts from axion stars moving
  through pulsar magnetospheres}},}\ }\href {\doibase
  10.1103/PhysRevD.103.043015} {\bibfield  {journal} {\bibinfo  {journal}
  {Phys. Rev. D}\ }\textbf {\bibinfo {volume} {103}},\ \bibinfo {pages}
  {043015} (\bibinfo {year} {2021})}\BibitemShut {NoStop}%
\bibitem [{\citenamefont {G\'enolini}\ \emph {et~al.}(2020)\citenamefont
  {G\'enolini}, \citenamefont {Serpico},\ and\ \citenamefont
  {Tinyakov}}]{Genolini:2020ejw}%
  \BibitemOpen
  \bibfield  {author} {\bibinfo {author} {\bibfnamefont {Y.}~\bibnamefont
  {G\'enolini}}, \bibinfo {author} {\bibfnamefont {P.}~\bibnamefont {Serpico}},
  \ and\ \bibinfo {author} {\bibfnamefont {P.}~\bibnamefont {Tinyakov}},\
  }\bibfield  {title} {\enquote {\bibinfo {title} {{Revisiting primordial black
  hole capture into neutron stars}},}\ }\href {\doibase
  10.1103/PhysRevD.102.083004} {\bibfield  {journal} {\bibinfo  {journal}
  {Phys. Rev. D}\ }\textbf {\bibinfo {volume} {102}},\ \bibinfo {pages}
  {083004} (\bibinfo {year} {2020})}\BibitemShut {NoStop}%
\bibitem [{\citenamefont {Kainulainen}\ \emph {et~al.}(2021)\citenamefont
  {Kainulainen}, \citenamefont {Nurmi}, \citenamefont {Schiappacasse},\ and\
  \citenamefont {Yanagida}}]{Kainulainen:2021rbg}%
  \BibitemOpen
  \bibfield  {author} {\bibinfo {author} {\bibfnamefont {Kimmo}\ \bibnamefont
  {Kainulainen}}, \bibinfo {author} {\bibfnamefont {Sami}\ \bibnamefont
  {Nurmi}}, \bibinfo {author} {\bibfnamefont {Enrico~D.}\ \bibnamefont
  {Schiappacasse}}, \ and\ \bibinfo {author} {\bibfnamefont {Tsutomu~T.}\
  \bibnamefont {Yanagida}},\ }\bibfield  {title} {\enquote {\bibinfo {title}
  {{Can primordial black holes as all dark matter explain fast radio
  bursts?}}}\ }\href {\doibase 10.1103/PhysRevD.104.123033} {\bibfield
  {journal} {\bibinfo  {journal} {Phys. Rev. D}\ }\textbf {\bibinfo {volume}
  {104}},\ \bibinfo {pages} {123033} (\bibinfo {year} {2021})}\BibitemShut
  {NoStop}%
\bibitem [{\citenamefont {Bhattacharjee}\ \emph {et~al.}(2013)\citenamefont
  {Bhattacharjee}, \citenamefont {Chaudhury}, \citenamefont {Kundu},\ and\
  \citenamefont {Majumdar}}]{Bhattacharjee:2012xm}%
  \BibitemOpen
  \bibfield  {author} {\bibinfo {author} {\bibfnamefont {Pijushpani}\
  \bibnamefont {Bhattacharjee}}, \bibinfo {author} {\bibfnamefont {Soumini}\
  \bibnamefont {Chaudhury}}, \bibinfo {author} {\bibfnamefont {Susmita}\
  \bibnamefont {Kundu}}, \ and\ \bibinfo {author} {\bibfnamefont {Subhabrata}\
  \bibnamefont {Majumdar}},\ }\bibfield  {title} {\enquote {\bibinfo {title}
  {{Sizing-up the WIMPs of Milky Way : Deriving the velocity distribution of
  Galactic Dark Matter particles from the rotation curve data}},}\ }\href
  {\doibase 10.1103/PhysRevD.87.083525} {\bibfield  {journal} {\bibinfo
  {journal} {Phys. Rev. D}\ }\textbf {\bibinfo {volume} {87}},\ \bibinfo
  {pages} {083525} (\bibinfo {year} {2013})}\BibitemShut {NoStop}%
\bibitem [{\citenamefont {{Navarro}}\ \emph {et~al.}(1996)\citenamefont
  {{Navarro}}, \citenamefont {{Frenk}},\ and\ \citenamefont
  {{White}}}]{1996ApJ...462..563N}%
  \BibitemOpen
  \bibfield  {author} {\bibinfo {author} {\bibfnamefont {Julio~F.}\
  \bibnamefont {{Navarro}}}, \bibinfo {author} {\bibfnamefont {Carlos~S.}\
  \bibnamefont {{Frenk}}}, \ and\ \bibinfo {author} {\bibfnamefont {Simon
  D.~M.}\ \bibnamefont {{White}}},\ }\bibfield  {title} {\enquote {\bibinfo
  {title} {{The Structure of Cold Dark Matter Halos}},}\ }\href {\doibase
  10.1086/177173} {\bibfield  {journal} {\bibinfo  {journal} {Astrophys. J.}\
  }\textbf {\bibinfo {volume} {462}},\ \bibinfo {pages} {563} (\bibinfo {year}
  {1996})}\BibitemShut {NoStop}%
\bibitem [{\citenamefont {Hertzberg}\ \emph {et~al.}(2021)\citenamefont
  {Hertzberg}, \citenamefont {Nurmi}, \citenamefont {Schiappacasse},\ and\
  \citenamefont {Yanagida}}]{Hertzberg:2020kpm}%
  \BibitemOpen
  \bibfield  {author} {\bibinfo {author} {\bibfnamefont {M.~P.}\ \bibnamefont
  {Hertzberg}}, \bibinfo {author} {\bibfnamefont {S.}~\bibnamefont {Nurmi}},
  \bibinfo {author} {\bibfnamefont {E.~D.}\ \bibnamefont {Schiappacasse}}, \
  and\ \bibinfo {author} {\bibfnamefont {T.~T.}\ \bibnamefont {Yanagida}},\
  }\bibfield  {title} {\enquote {\bibinfo {title} {{Shining Primordial Black
  Holes}},}\ }\href {\doibase 10.1103/PhysRevD.103.063025} {\bibfield
  {journal} {\bibinfo  {journal} {Phys. Rev. D}\ }\textbf {\bibinfo {volume}
  {103}},\ \bibinfo {pages} {063025} (\bibinfo {year} {2021})}\BibitemShut
  {NoStop}%
\bibitem [{\citenamefont {{Freudenreich}}(1998)}]{1998ApJ...492..495F}%
  \BibitemOpen
  \bibfield  {author} {\bibinfo {author} {\bibfnamefont {H.~T.}\ \bibnamefont
  {{Freudenreich}}},\ }\bibfield  {title} {\enquote {\bibinfo {title} {{A COBE
  Model of the Galactic Bar and Disk}},}\ }\href {\doibase 10.1086/305065}
  {\bibfield  {journal} {\bibinfo  {journal} {Astrophys. J.}\ }\textbf
  {\bibinfo {volume} {492}},\ \bibinfo {pages} {495} (\bibinfo {year}
  {1998})}\BibitemShut {NoStop}%
\bibitem [{\citenamefont {Catena}\ and\ \citenamefont
  {Ullio}(2012)}]{Catena:2011kv}%
  \BibitemOpen
  \bibfield  {author} {\bibinfo {author} {\bibfnamefont {Riccardo}\
  \bibnamefont {Catena}}\ and\ \bibinfo {author} {\bibfnamefont {Piero}\
  \bibnamefont {Ullio}},\ }\bibfield  {title} {\enquote {\bibinfo {title} {{The
  local dark matter phase-space density and impact on WIMP direct
  detection}},}\ }\href {\doibase 10.1088/1475-7516/2012/05/005} {\bibfield
  {journal} {\bibinfo  {journal} {J. Cosmol. Astropart. Phys.}\ }\textbf
  {\bibinfo {volume} {05}},\ \bibinfo {pages} {005} (\bibinfo {year}
  {2012})}\BibitemShut {NoStop}%
\bibitem [{\citenamefont {{Eddington}}(1916)}]{1916MNRAS..76..572E}%
  \BibitemOpen
  \bibfield  {author} {\bibinfo {author} {\bibfnamefont {A.~S.}\ \bibnamefont
  {{Eddington}}},\ }\bibfield  {title} {\enquote {\bibinfo {title} {{The
  distribution of stars in globular clusters}},}\ }\href {\doibase
  "10.1093/mnras/76.7.572"} {\bibfield  {journal} {\bibinfo  {journal} {Mon.
  Not. R. Astron. Soc.}\ }\textbf {\bibinfo {volume} {76}},\ \bibinfo {pages}
  {572} (\bibinfo {year} {1916})}\BibitemShut {NoStop}%
\bibitem [{\citenamefont {{Binney}}\ and\ \citenamefont
  {{Tremaine}}(2008)}]{2008gady.book.....B}%
  \BibitemOpen
  \bibfield  {author} {\bibinfo {author} {\bibfnamefont {James}\ \bibnamefont
  {{Binney}}}\ and\ \bibinfo {author} {\bibfnamefont {Scott}\ \bibnamefont
  {{Tremaine}}},\ }\href@noop {} {\emph {\bibinfo {title} {{Galactic Dynamics:
  Second Edition}}}}\ (\bibinfo  {publisher} {Princeton University Press, NJ,
  USA},\ \bibinfo {year} {2008})\BibitemShut {NoStop}%
\bibitem [{\citenamefont {Kormendy}\ and\ \citenamefont
  {Ho}()}]{Kormendy:2000cf}%
  \BibitemOpen
  \bibfield  {author} {\bibinfo {author} {\bibfnamefont {John}\ \bibnamefont
  {Kormendy}}\ and\ \bibinfo {author} {\bibfnamefont {Luis~C.}\ \bibnamefont
  {Ho}},\ }\bibfield  {title} {\enquote {\bibinfo {title} {{Supermassive black
  holes in inactive galaxies}},}\ }\href@noop {} {\bibinfo  {journal}
  {\textit{Encyclopedia of Astronomy} \& \textit{Astrophysics} (CRC Press, Boca
  Raton, 2001)}\ }\BibitemShut {NoStop}%
\bibitem [{\citenamefont {{Peebles}}(1972)}]{1972ApJ...178..371P}%
  \BibitemOpen
\bibfield  {journal} {  }\bibfield  {author} {\bibinfo {author} {\bibfnamefont
  {P.~J.~E.}\ \bibnamefont {{Peebles}}},\ }\bibfield  {title} {\enquote
  {\bibinfo {title} {{Star Distribution Near a Collapsed Object}},}\ }\href
  {\doibase 10.1086/151797} {\bibfield  {journal} {\bibinfo  {journal}
  {Astrophys. J.}\ }\textbf {\bibinfo {volume} {178}},\ \bibinfo {pages} {371}
  (\bibinfo {year} {1972})}\BibitemShut {NoStop}%
\bibitem [{\citenamefont {Safdi}\ \emph {et~al.}(2019)\citenamefont {Safdi},
  \citenamefont {Sun},\ and\ \citenamefont {Chen}}]{Safdi:2018oeu}%
  \BibitemOpen
  \bibfield  {author} {\bibinfo {author} {\bibfnamefont {Benjamin~R.}\
  \bibnamefont {Safdi}}, \bibinfo {author} {\bibfnamefont {Zhiquan}\
  \bibnamefont {Sun}}, \ and\ \bibinfo {author} {\bibfnamefont {Alexander~Y.}\
  \bibnamefont {Chen}},\ }\bibfield  {title} {\enquote {\bibinfo {title}
  {{Detecting Axion Dark Matter with Radio Lines from Neutron Star
  Populations}},}\ }\href {\doibase 10.1103/PhysRevD.99.123021} {\bibfield
  {journal} {\bibinfo  {journal} {Phys. Rev. D}\ }\textbf {\bibinfo {volume}
  {99}},\ \bibinfo {pages} {123021} (\bibinfo {year} {2019})}\BibitemShut
  {NoStop}%
\bibitem [{\citenamefont {Dehnen}\ \emph {et~al.}(2006)\citenamefont {Dehnen},
  \citenamefont {McLaughlin},\ and\ \citenamefont {Sachania}}]{Dehnen:2006cm}%
  \BibitemOpen
  \bibfield  {author} {\bibinfo {author} {\bibfnamefont {Walter}\ \bibnamefont
  {Dehnen}}, \bibinfo {author} {\bibfnamefont {Dean}\ \bibnamefont
  {McLaughlin}}, \ and\ \bibinfo {author} {\bibfnamefont {Jalpesh}\
  \bibnamefont {Sachania}},\ }\bibfield  {title} {\enquote {\bibinfo {title}
  {{The velocity dispersion and mass profile of the milky way}},}\ }\href
  {\doibase 10.1111/j.1365-2966.2006.10404.x} {\bibfield  {journal} {\bibinfo
  {journal} {Mon. Not. R. Astron. Soc.}\ }\textbf {\bibinfo {volume} {369}},\
  \bibinfo {pages} {1688} (\bibinfo {year} {2006})}\BibitemShut {NoStop}%
\bibitem [{\citenamefont {Sasaki~\textit{et al.}}(2018)}]{Sasaki:2018dmp}%
  \BibitemOpen
  \bibfield  {author} {\bibinfo {author} {\bibfnamefont {M.}~\bibnamefont
  {Sasaki~\textit{et al.}}},\ }\bibfield  {title} {\enquote {\bibinfo {title}
  {{Primordial black holes\textemdash{}perspectives in gravitational wave
  astronomy}},}\ }\href {\doibase 10.1088/1361-6382/aaa7b4} {\bibfield
  {journal} {\bibinfo  {journal} {Classical Quantum Gravity}\ }\textbf
  {\bibinfo {volume} {35}},\ \bibinfo {pages} {063001} (\bibinfo {year}
  {2018})}\BibitemShut {NoStop}%
\bibitem [{\citenamefont {Kormendy}\ and\ \citenamefont
  {Richstone}(1995)}]{Kormendy:1995er}%
  \BibitemOpen
  \bibfield  {author} {\bibinfo {author} {\bibfnamefont {John}\ \bibnamefont
  {Kormendy}}\ and\ \bibinfo {author} {\bibfnamefont {Douglas}\ \bibnamefont
  {Richstone}},\ }\bibfield  {title} {\enquote {\bibinfo {title} {{Inward
  bound: The Search for supermassive black holes in galactic nuclei}},}\ }\href
  {\doibase 10.1146/annurev.aa.33.090195.003053} {\bibfield  {journal}
  {\bibinfo  {journal} {Ann. Rev. Astron. Astrophys.}\ }\textbf {\bibinfo
  {volume} {33}},\ \bibinfo {pages} {581} (\bibinfo {year} {1995})}\BibitemShut
  {NoStop}%
\bibitem [{\citenamefont {Kormendy}\ and\ \citenamefont
  {Ho}(2013)}]{Kormendy:2013dxa}%
  \BibitemOpen
  \bibfield  {author} {\bibinfo {author} {\bibfnamefont {John}\ \bibnamefont
  {Kormendy}}\ and\ \bibinfo {author} {\bibfnamefont {Luis~C.}\ \bibnamefont
  {Ho}},\ }\bibfield  {title} {\enquote {\bibinfo {title} {{Coevolution (Or
  Not) of Supermassive Black Holes and Host Galaxies}},}\ }\href {\doibase
  10.1146/annurev-astro-082708-101811} {\bibfield  {journal} {\bibinfo
  {journal} {Ann. Rev. Astron. Astrophys.}\ }\textbf {\bibinfo {volume} {51}},\
  \bibinfo {pages} {511} (\bibinfo {year} {2013})}\BibitemShut {NoStop}%
\bibitem [{\citenamefont {{Bahcall}}\ and\ \citenamefont
  {{Wolf}}(1976)}]{1976ApJ...209..214B}%
  \BibitemOpen
  \bibfield  {author} {\bibinfo {author} {\bibfnamefont {J.~N.}\ \bibnamefont
  {{Bahcall}}}\ and\ \bibinfo {author} {\bibfnamefont {R.~A.}\ \bibnamefont
  {{Wolf}}},\ }\bibfield  {title} {\enquote {\bibinfo {title} {{Star
  distribution around a massive black hole in a globular cluster.}}}\ }\href
  {\doibase 10.1086/154711} {\bibfield  {journal} {\bibinfo  {journal}
  {Astrophys. J.}\ }\textbf {\bibinfo {volume} {209}},\ \bibinfo {pages} {214}
  (\bibinfo {year} {1976})}\BibitemShut {NoStop}%
\bibitem [{\citenamefont {Freitag}\ \emph {et~al.}(2006)\citenamefont
  {Freitag}, \citenamefont {Amaro-Seoane},\ and\ \citenamefont
  {Kalogera}}]{Freitag:2006qf}%
  \BibitemOpen
  \bibfield  {author} {\bibinfo {author} {\bibfnamefont {Marc}\ \bibnamefont
  {Freitag}}, \bibinfo {author} {\bibfnamefont {Pau}\ \bibnamefont
  {Amaro-Seoane}}, \ and\ \bibinfo {author} {\bibfnamefont {Vassiliki}\
  \bibnamefont {Kalogera}},\ }\bibfield  {title} {\enquote {\bibinfo {title}
  {{Stellar remnants in galactic nuclei: mass segregation}},}\ }\href {\doibase
  10.1086/506193} {\bibfield  {journal} {\bibinfo  {journal} {Astrophys. J.}\
  }\textbf {\bibinfo {volume} {649}},\ \bibinfo {pages} {91} (\bibinfo {year}
  {2006})}\BibitemShut {NoStop}%
\bibitem [{\citenamefont {{Dehnen}}(1993)}]{1993MNRAS.265..250D}%
  \BibitemOpen
  \bibfield  {author} {\bibinfo {author} {\bibfnamefont {W.}~\bibnamefont
  {{Dehnen}}},\ }\bibfield  {title} {\enquote {\bibinfo {title} {{A Family of
  Potential-Density Pairs for Spherical Galaxies and Bulges}},}\ }\href
  {\doibase "10.1093/mnras/265.1.250"} {\bibfield  {journal} {\bibinfo
  {journal} {Mon. Not. R. Astron. Soc.}\ }\textbf {\bibinfo {volume} {265}},\
  \bibinfo {pages} {250} (\bibinfo {year} {1993})}\BibitemShut {NoStop}%
\bibitem [{\citenamefont {{Tremaine}}\ \emph {et~al.}(1994)\citenamefont
  {{Tremaine}}, \citenamefont {{Richstone}}, \citenamefont {{Byun}},
  \citenamefont {{Dressler}}, \citenamefont {{Faber}}, \citenamefont
  {{Grillmair}}, \citenamefont {{Kormendy}},\ and\ \citenamefont
  {{Lauer}}}]{1994AJ....107..634T}%
  \BibitemOpen
  \bibfield  {author} {\bibinfo {author} {\bibfnamefont {Scott}\ \bibnamefont
  {{Tremaine}}}, \bibinfo {author} {\bibfnamefont {Douglas~O.}\ \bibnamefont
  {{Richstone}}}, \bibinfo {author} {\bibfnamefont {Yong-Ik}\ \bibnamefont
  {{Byun}}}, \bibinfo {author} {\bibfnamefont {Alan}\ \bibnamefont
  {{Dressler}}}, \bibinfo {author} {\bibfnamefont {S.~M.}\ \bibnamefont
  {{Faber}}}, \bibinfo {author} {\bibfnamefont {Carl}\ \bibnamefont
  {{Grillmair}}}, \bibinfo {author} {\bibfnamefont {John}\ \bibnamefont
  {{Kormendy}}}, \ and\ \bibinfo {author} {\bibfnamefont {Tod~R.}\ \bibnamefont
  {{Lauer}}},\ }\bibfield  {title} {\enquote {\bibinfo {title} {{A Family of
  Models for Spherical Stellar Systems}},}\ }\href {\doibase 10.1086/116883}
  {\bibfield  {journal} {\bibinfo  {journal} {Astron. J.}\ }\textbf {\bibinfo
  {volume} {107}},\ \bibinfo {pages} {634} (\bibinfo {year}
  {1994})}\BibitemShut {NoStop}%
\bibitem [{\citenamefont {{Ilyin}}\ \emph {et~al.}(2004)\citenamefont
  {{Ilyin}}, \citenamefont {{Zybin}},\ and\ \citenamefont
  {{Gurevich}}}]{2004JETP...98....1I}%
  \BibitemOpen
  \bibfield  {author} {\bibinfo {author} {\bibfnamefont {A.~S.}\ \bibnamefont
  {{Ilyin}}}, \bibinfo {author} {\bibfnamefont {K.~P.}\ \bibnamefont
  {{Zybin}}}, \ and\ \bibinfo {author} {\bibfnamefont {A.~V.}\ \bibnamefont
  {{Gurevich}}},\ }\bibfield  {title} {\enquote {\bibinfo {title} {{Dark matter
  in galaxies and the growth of giant black holes}},}\ }\href {\doibase
  10.1134/1.1648097} {\bibfield  {journal} {\bibinfo  {journal} {Sov. J. Exp.
  Theor. Phys.}\ }\textbf {\bibinfo {volume} {98}},\ \bibinfo {pages} {1}
  (\bibinfo {year} {2004})}\BibitemShut {NoStop}%
\bibitem [{\citenamefont {Merritt}(2004)}]{PhysRevLett.92.201304}%
  \BibitemOpen
  \bibfield  {author} {\bibinfo {author} {\bibfnamefont {David}\ \bibnamefont
  {Merritt}},\ }\bibfield  {title} {\enquote {\bibinfo {title} {Evolution of
  the dark matter distribution at the admx},}\ }\href {\doibase
  10.1103/PhysRevLett.92.201304} {\bibfield  {journal} {\bibinfo  {journal}
  {Phys. Rev. Lett.}\ }\textbf {\bibinfo {volume} {92}},\ \bibinfo {pages}
  {201304} (\bibinfo {year} {2004})}\BibitemShut {NoStop}%
\bibitem [{\citenamefont {Gnedin}\ and\ \citenamefont
  {Primack}(2004)}]{PhysRevLett.93.061302}%
  \BibitemOpen
  \bibfield  {author} {\bibinfo {author} {\bibfnamefont {Oleg~Y.}\ \bibnamefont
  {Gnedin}}\ and\ \bibinfo {author} {\bibfnamefont {Joel~R.}\ \bibnamefont
  {Primack}},\ }\bibfield  {title} {\enquote {\bibinfo {title} {Dark matter
  profile in the admx},}\ }\href {\doibase 10.1103/PhysRevLett.93.061302}
  {\bibfield  {journal} {\bibinfo  {journal} {Phys. Rev. Lett.}\ }\textbf
  {\bibinfo {volume} {93}},\ \bibinfo {pages} {061302} (\bibinfo {year}
  {2004})}\BibitemShut {NoStop}%
\bibitem [{\citenamefont {Merritt}\ \emph {et~al.}(2007)\citenamefont
  {Merritt}, \citenamefont {Harfst},\ and\ \citenamefont
  {Bertone}}]{Merritt:2006mt}%
  \BibitemOpen
  \bibfield  {author} {\bibinfo {author} {\bibfnamefont {David}\ \bibnamefont
  {Merritt}}, \bibinfo {author} {\bibfnamefont {Stefan}\ \bibnamefont
  {Harfst}}, \ and\ \bibinfo {author} {\bibfnamefont {Gianfranco}\ \bibnamefont
  {Bertone}},\ }\bibfield  {title} {\enquote {\bibinfo {title} {{Collisionally
  Regenerated Dark Matter Structures in Galactic Nuclei}},}\ }\href {\doibase
  10.1103/PhysRevD.75.043517} {\bibfield  {journal} {\bibinfo  {journal} {Phys.
  Rev. D}\ }\textbf {\bibinfo {volume} {75}},\ \bibinfo {pages} {043517}
  (\bibinfo {year} {2007})}\BibitemShut {NoStop}%
\bibitem [{\citenamefont {Navarro}\ \emph {et~al.}(1996)\citenamefont
  {Navarro}, \citenamefont {Frenk},\ and\ \citenamefont
  {White}}]{Navarro:1995iw}%
  \BibitemOpen
  \bibfield  {author} {\bibinfo {author} {\bibfnamefont {Julio~F.}\
  \bibnamefont {Navarro}}, \bibinfo {author} {\bibfnamefont {Carlos~S.}\
  \bibnamefont {Frenk}}, \ and\ \bibinfo {author} {\bibfnamefont {Simon D.~M.}\
  \bibnamefont {White}},\ }\bibfield  {title} {\enquote {\bibinfo {title} {{The
  Structure of cold dark matter halos}},}\ }\href {\doibase 10.1086/177173}
  {\bibfield  {journal} {\bibinfo  {journal} {Astrophys. J.}\ }\textbf
  {\bibinfo {volume} {462}},\ \bibinfo {pages} {563} (\bibinfo {year}
  {1996})}\BibitemShut {NoStop}%
\bibitem [{\citenamefont {{Moore}}\ \emph {et~al.}(1998)\citenamefont
  {{Moore}}, \citenamefont {{Governato}}, \citenamefont {{Quinn}},
  \citenamefont {{Stadel}},\ and\ \citenamefont
  {{Lake}}}]{1998ApJ...499L...5M}%
  \BibitemOpen
  \bibfield  {author} {\bibinfo {author} {\bibfnamefont {B.}~\bibnamefont
  {{Moore}}}, \bibinfo {author} {\bibfnamefont {F.}~\bibnamefont
  {{Governato}}}, \bibinfo {author} {\bibfnamefont {T.}~\bibnamefont
  {{Quinn}}}, \bibinfo {author} {\bibfnamefont {J.}~\bibnamefont {{Stadel}}}, \
  and\ \bibinfo {author} {\bibfnamefont {G.}~\bibnamefont {{Lake}}},\
  }\bibfield  {title} {\enquote {\bibinfo {title} {{Resolving the Structure of
  Cold Dark Matter Halos}},}\ }\href {\doibase 10.1086/311333} {\bibfield
  {journal} {\bibinfo  {journal} {Astrophys. J. Lett.}\ }\textbf {\bibinfo
  {volume} {499}},\ \bibinfo {pages} {L5} (\bibinfo {year} {1998})}\BibitemShut
  {NoStop}%
\bibitem [{\citenamefont {Gondolo}\ and\ \citenamefont
  {Silk}(1999)}]{Gondolo:1999ef}%
  \BibitemOpen
  \bibfield  {author} {\bibinfo {author} {\bibfnamefont {Paolo}\ \bibnamefont
  {Gondolo}}\ and\ \bibinfo {author} {\bibfnamefont {Joseph}\ \bibnamefont
  {Silk}},\ }\bibfield  {title} {\enquote {\bibinfo {title} {{Dark matter
  annihilation at the galactic center}},}\ }\href {\doibase
  10.1103/PhysRevLett.83.1719} {\bibfield  {journal} {\bibinfo  {journal}
  {Phys. Rev. Lett.}\ }\textbf {\bibinfo {volume} {83}},\ \bibinfo {pages}
  {1719} (\bibinfo {year} {1999})}\BibitemShut {NoStop}%
\bibitem [{\citenamefont {Bertone}\ \emph {et~al.}(2002)\citenamefont
  {Bertone}, \citenamefont {Sigl},\ and\ \citenamefont
  {Silk}}]{Bertone:2002je}%
  \BibitemOpen
  \bibfield  {author} {\bibinfo {author} {\bibfnamefont {Gianfranco}\
  \bibnamefont {Bertone}}, \bibinfo {author} {\bibfnamefont {Guenter}\
  \bibnamefont {Sigl}}, \ and\ \bibinfo {author} {\bibfnamefont {Joseph}\
  \bibnamefont {Silk}},\ }\bibfield  {title} {\enquote {\bibinfo {title}
  {{Annihilation radiation from a dark matter spike at the galactic center}},}\
  }\href {\doibase 10.1046/j.1365-8711.2002.05892.x} {\bibfield  {journal}
  {\bibinfo  {journal} {Mon. Not. R. Astron. Soc.}\ }\textbf {\bibinfo {volume}
  {337}},\ \bibinfo {pages} {98} (\bibinfo {year} {2002})}\BibitemShut
  {NoStop}%
\bibitem [{\citenamefont {Ullio}\ \emph {et~al.}(2001)\citenamefont {Ullio},
  \citenamefont {Zhao},\ and\ \citenamefont {Kamionkowski}}]{Ullio:2001fb}%
  \BibitemOpen
  \bibfield  {author} {\bibinfo {author} {\bibfnamefont {Piero}\ \bibnamefont
  {Ullio}}, \bibinfo {author} {\bibfnamefont {HongSheng}\ \bibnamefont {Zhao}},
  \ and\ \bibinfo {author} {\bibfnamefont {Marc}\ \bibnamefont
  {Kamionkowski}},\ }\bibfield  {title} {\enquote {\bibinfo {title} {{A Dark
  matter spike at the galactic center?}}}\ }\href {\doibase
  10.1103/PhysRevD.64.043504} {\bibfield  {journal} {\bibinfo  {journal} {Phys.
  Rev. D}\ }\textbf {\bibinfo {volume} {64}},\ \bibinfo {pages} {043504}
  (\bibinfo {year} {2001})}\BibitemShut {NoStop}%
\bibitem [{\citenamefont {Nurmi}\ \emph {et~al.}(2021)\citenamefont {Nurmi},
  \citenamefont {Schiappacasse},\ and\ \citenamefont
  {Yanagida}}]{Nurmi:2021xds}%
  \BibitemOpen
  \bibfield  {author} {\bibinfo {author} {\bibfnamefont {S.}~\bibnamefont
  {Nurmi}}, \bibinfo {author} {\bibfnamefont {E.~D.}\ \bibnamefont
  {Schiappacasse}}, \ and\ \bibinfo {author} {\bibfnamefont {T.~T.}\
  \bibnamefont {Yanagida}},\ }\bibfield  {title} {\enquote {\bibinfo {title}
  {{Radio signatures from encounters between neutron stars and QCD-axion
  minihalos around primordial black~holes}},}\ }\href {\doibase
  10.1088/1475-7516/2021/09/004} {\bibfield  {journal} {\bibinfo  {journal} {J.
  Cosmol. Astropart. Phys.}\ }\textbf {\bibinfo {volume} {09}},\ \bibinfo
  {pages} {004} (\bibinfo {year} {2021})}\BibitemShut {NoStop}%
\bibitem [{\citenamefont {Champion~\textit{et al.}}(2016)}]{Champion:2015pmj}%
  \BibitemOpen
  \bibfield  {author} {\bibinfo {author} {\bibfnamefont {D.~J.}\ \bibnamefont
  {Champion~\textit{et al.}}},\ }\bibfield  {title} {\enquote {\bibinfo {title}
  {{Fice new Fast Radio Bursts from the HTRU high-latitude survey at Parkes:
  first evidence for two-component bursts}},}\ }\href {\doibase
  10.1093/mnrasl/slw069} {\bibfield  {journal} {\bibinfo  {journal} {Mon. Not.
  R. Astron. Soc.}\ }\textbf {\bibinfo {volume} {460}},\ \bibinfo {pages} {L30}
  (\bibinfo {year} {2016})},\ \Eprint {http://arxiv.org/abs/1511.07746}
  {arXiv:1511.07746 [astro-ph.HE]} \BibitemShut {NoStop}%
\bibitem [{\citenamefont {Petroff}\ \emph {et~al.}(2019)\citenamefont
  {Petroff}, \citenamefont {Hessels},\ and\ \citenamefont
  {Lorimer}}]{Petroff:2019tty}%
  \BibitemOpen
  \bibfield  {author} {\bibinfo {author} {\bibfnamefont {E.}~\bibnamefont
  {Petroff}}, \bibinfo {author} {\bibfnamefont {J.~W.~T.}\ \bibnamefont
  {Hessels}}, \ and\ \bibinfo {author} {\bibfnamefont {D.~R.}\ \bibnamefont
  {Lorimer}},\ }\bibfield  {title} {\enquote {\bibinfo {title} {{Fast Radio
  Bursts}},}\ }\href {\doibase 10.1007/s00159-019-0116-6} {\bibfield  {journal}
  {\bibinfo  {journal} {Astron. Astrophys. Rev.}\ }\textbf {\bibinfo {volume}
  {27}},\ \bibinfo {pages} {4} (\bibinfo {year} {2019})}\BibitemShut {NoStop}%
\bibitem [{\citenamefont {Piratova-Moreno}\ and\ \citenamefont
  {Garc\'ia}(2024)}]{Piratova:2024}%
  \BibitemOpen
  \bibfield  {author} {\bibinfo {author} {\bibfnamefont {E.~F.}\ \bibnamefont
  {Piratova-Moreno}}\ and\ \bibinfo {author} {\bibfnamefont {L.~A.}\
  \bibnamefont {Garc\'ia}},\ }\bibfield  {title} {\enquote {\bibinfo {title}
  {{Modeling the dispersion measure—redshift relation for fast radio
  bursts}},}\ }\href {\doibase 10.3389/fspas.2024.1371787} {\bibfield
  {journal} {\bibinfo  {journal} {Front. Astron. Space Sci.}\ }\textbf
  {\bibinfo {volume} {11}} (\bibinfo {year} {2024}),\
  10.3389/fspas.2024.1371787}\BibitemShut {NoStop}%
\bibitem [{\citenamefont {Goodwin}\ \emph {et~al.}(1997)\citenamefont
  {Goodwin}, \citenamefont {Gribbin},\ and\ \citenamefont
  {Hendry}}]{Goodwin:1997ys}%
  \BibitemOpen
  \bibfield  {author} {\bibinfo {author} {\bibfnamefont {S.~P.}\ \bibnamefont
  {Goodwin}}, \bibinfo {author} {\bibfnamefont {J.}~\bibnamefont {Gribbin}}, \
  and\ \bibinfo {author} {\bibfnamefont {M.~A.}\ \bibnamefont {Hendry}},\
  }\bibfield  {title} {\enquote {\bibinfo {title} {{The milky way is just an
  average spiral}},}\ }\href@noop {} {\  (\bibinfo {year} {1997})},\ \Eprint
  {http://arxiv.org/abs/astro-ph/9704216} {arXiv:astro-ph/9704216} \BibitemShut
  {NoStop}%
\bibitem [{\citenamefont {Cirelli}\ \emph {et~al.}(2024)\citenamefont
  {Cirelli}, \citenamefont {Strumia},\ and\ \citenamefont
  {Zupan}}]{Cirelli:2024ssz}%
  \BibitemOpen
  \bibfield  {author} {\bibinfo {author} {\bibfnamefont {Marco}\ \bibnamefont
  {Cirelli}}, \bibinfo {author} {\bibfnamefont {Alessandro}\ \bibnamefont
  {Strumia}}, \ and\ \bibinfo {author} {\bibfnamefont {Jure}\ \bibnamefont
  {Zupan}},\ }\bibfield  {title} {\enquote {\bibinfo {title} {{Dark Matter}},}\
  }\href@noop {} {\  (\bibinfo {year} {2024})},\ \Eprint
  {http://arxiv.org/abs/2406.01705} {arXiv:2406.01705 [hep-ph]} \BibitemShut
  {NoStop}%
\bibitem [{\citenamefont {{Sofue}}(2017)}]{2017PASJ...69R...1S}%
  \BibitemOpen
  \bibfield  {author} {\bibinfo {author} {\bibfnamefont {Yoshiaki}\
  \bibnamefont {{Sofue}}},\ }\bibfield  {title} {\enquote {\bibinfo {title}
  {{Rotation and mass in the Milky Way and spiral galaxies}},}\ }\href
  {\doibase 10.1093/pasj/psw103} {\bibfield  {journal} {\bibinfo  {journal}
  {Publ. Astron. Soc. Jpn.}\ }\textbf {\bibinfo {volume} {69}},\ \bibinfo {eid}
  {R1} (\bibinfo {year} {2017})}\BibitemShut {NoStop}%
\bibitem [{\citenamefont {{{Crnojevi{\'c}}, D., {Mutlu-Pakdil},
  B.}}(2021)}]{Crnoj:2021}%
  \BibitemOpen
  \bibfield  {author} {\bibinfo {author} {\bibnamefont {{{Crnojevi{\'c}}, D.,
  {Mutlu-Pakdil}, B.}}},\ }\bibfield  {title} {\enquote {\bibinfo {title}
  {{Dwarf galaxies yesterday, now and tomorrow}},}\ }\href {\doibase
  10.1038/s41550-021-01563-1} {\bibfield  {journal} {\bibinfo  {journal} {Nat.
  Astron.}\ ,\ \bibinfo {pages} {1191}} (\bibinfo {year} {2021})}\BibitemShut
  {NoStop}%
\bibitem [{\citenamefont {{Sellwood}}\ and\ \citenamefont
  {{Masters}}(2022)}]{2022ARA&A..60...73S}%
  \BibitemOpen
  \bibfield  {author} {\bibinfo {author} {\bibfnamefont {J.~A.}\ \bibnamefont
  {{Sellwood}}}\ and\ \bibinfo {author} {\bibfnamefont {Karen~L.}\ \bibnamefont
  {{Masters}}},\ }\bibfield  {title} {\enquote {\bibinfo {title} {{Spirals in
  Galaxies}},}\ }\href {\doibase 10.1146/annurev-astro-052920-104505}
  {\bibfield  {journal} {\bibinfo  {journal} {Ann. Rev. Astron. Astrophys.}\
  }\textbf {\bibinfo {volume} {60}} (\bibinfo {year} {2022}),\
  10.1146/annurev-astro-052920-104505}\BibitemShut {NoStop}%
\bibitem [{\citenamefont {{Annibali, F.}}(2022)}]{2022Annibali}%
  \BibitemOpen
  \bibfield  {author} {\bibinfo {author} {\bibfnamefont {{Tosi, M.}}\
  \bibnamefont {{Annibali, F.}}},\ }\bibfield  {title} {\enquote {\bibinfo
  {title} {{Chemical and stellar properties of star-forming dwarf galaxies}},}\
  }\href {\doibase 10.1038/s41550-021-01575-x} {\bibfield  {journal} {\bibinfo
  {journal} {Nat. Astron.}\ }\textbf {\bibinfo {volume} {6}},\ \bibinfo {pages}
  {48} (\bibinfo {year} {2022})}\BibitemShut {NoStop}%
\bibitem [{\citenamefont {Loveday}(1996)}]{Loveday:1996xn}%
  \BibitemOpen
  \bibfield  {author} {\bibinfo {author} {\bibfnamefont {J.}~\bibnamefont
  {Loveday}},\ }\bibfield  {title} {\enquote {\bibinfo {title} {{The APM bright
  galaxy catalog}},}\ }\href {\doibase 10.1093/mnras/278.4.1025} {\bibfield
  {journal} {\bibinfo  {journal} {Mon. Not. R. Astron. Soc.}\ }\textbf
  {\bibinfo {volume} {278}},\ \bibinfo {pages} {1025} (\bibinfo {year}
  {1996})}\BibitemShut {NoStop}%
\bibitem [{\citenamefont {{Bechtol \textit{et
  al.}}}(2015)}]{2015ApJ...807...50B}%
  \BibitemOpen
  \bibfield  {author} {\bibinfo {author} {\bibfnamefont {K.}~\bibnamefont
  {{Bechtol \textit{et al.}}}},\ }\bibfield  {title} {\enquote {\bibinfo
  {title} {{Eight New Milky Way Companions Discovered in First-year Dark Energy
  Survey Data}},}\ }\href {\doibase 10.1088/0004-637X/807/1/50} {\bibfield
  {journal} {\bibinfo  {journal} {\apj}\ }\textbf {\bibinfo {volume} {807}},\
  \bibinfo {eid} {50} (\bibinfo {year} {2015})}\BibitemShut {NoStop}%
\bibitem [{\citenamefont {{Drlica-Wagner \textit{et
  al.}}}(2015)}]{2015ApJ...813..109D}%
  \BibitemOpen
  \bibfield  {author} {\bibinfo {author} {\bibfnamefont {A.}~\bibnamefont
  {{Drlica-Wagner \textit{et al.}}}},\ }\bibfield  {title} {\enquote {\bibinfo
  {title} {{Eight Ultra-faint Galaxy Candidates Discovered in Year Two of the
  Dark Energy Survey}},}\ }\href {\doibase 10.1088/0004-637X/813/2/109}
  {\bibfield  {journal} {\bibinfo  {journal} {\apj}\ }\textbf {\bibinfo
  {volume} {813}},\ \bibinfo {eid} {109} (\bibinfo {year} {2015})}\BibitemShut
  {NoStop}%
\bibitem [{\citenamefont {{Martin \textit{et
  al.}}}(2015)}]{2015ApJ...804L...5M}%
  \BibitemOpen
  \bibfield  {author} {\bibinfo {author} {\bibfnamefont {Nicolas~F.}\
  \bibnamefont {{Martin \textit{et al.}}}},\ }\bibfield  {title} {\enquote
  {\bibinfo {title} {{Hydra II: A Faint and Compact Milky Way Dwarf Galaxy
  Found in the Survey of the Magellanic Stellar History}},}\ }\href {\doibase
  10.1088/2041-8205/804/1/L5} {\bibfield  {journal} {\bibinfo  {journal}
  {Astrophys. J. Lett.}\ }\textbf {\bibinfo {volume} {804}},\ \bibinfo {eid}
  {L5} (\bibinfo {year} {2015})},\ \Eprint {http://arxiv.org/abs/1503.06216}
  {arXiv:1503.06216 [astro-ph.GA]} \BibitemShut {NoStop}%
\bibitem [{\citenamefont {Cerny~\textit{et al.}}(2021)}]{2021ApJ...920L..44C}%
  \BibitemOpen
  \bibfield  {author} {\bibinfo {author} {\bibfnamefont {W.}~\bibnamefont
  {Cerny~\textit{et al.}}},\ }\bibfield  {title} {\enquote {\bibinfo {title}
  {{Eridanus IV: An ultra-faint dwarf galaxy candidate discovered in the DECam
  local volume exploration survey}},}\ }\href {\doibase
  10.3847/2041-8213/ac2d9a} {\bibfield  {journal} {\bibinfo  {journal}
  {Astrophys. J. Lett.}\ }\textbf {\bibinfo {volume} {920}},\ \bibinfo {eid}
  {L44} (\bibinfo {year} {2021})}\BibitemShut {NoStop}%
\bibitem [{\citenamefont {{Torrealba}}\ \emph {et~al.}(2016)\citenamefont
  {{Torrealba}}, \citenamefont {{Koposov}}, \citenamefont {{Belokurov}},
  \citenamefont {{Irwin}}, \citenamefont {{Collins}}, \citenamefont
  {{Spencer}}, \citenamefont {{Ibata}}, \citenamefont {{Mateo}}, \citenamefont
  {{Bonaca}},\ and\ \citenamefont {{Jethwa}}}]{2016MNRAS.463..712T}%
  \BibitemOpen
  \bibfield  {author} {\bibinfo {author} {\bibfnamefont {G.}~\bibnamefont
  {{Torrealba}}}, \bibinfo {author} {\bibfnamefont {S.~E.}\ \bibnamefont
  {{Koposov}}}, \bibinfo {author} {\bibfnamefont {V.}~\bibnamefont
  {{Belokurov}}}, \bibinfo {author} {\bibfnamefont {M.}~\bibnamefont
  {{Irwin}}}, \bibinfo {author} {\bibfnamefont {M.}~\bibnamefont {{Collins}}},
  \bibinfo {author} {\bibfnamefont {M.}~\bibnamefont {{Spencer}}}, \bibinfo
  {author} {\bibfnamefont {R.}~\bibnamefont {{Ibata}}}, \bibinfo {author}
  {\bibfnamefont {M.}~\bibnamefont {{Mateo}}}, \bibinfo {author} {\bibfnamefont
  {A.}~\bibnamefont {{Bonaca}}}, \ and\ \bibinfo {author} {\bibfnamefont
  {P.}~\bibnamefont {{Jethwa}}},\ }\bibfield  {title} {\enquote {\bibinfo
  {title} {{At the survey limits: discovery of the Aquarius 2 dwarf galaxy in
  the VST ATLAS and the SDSS data}},}\ }\href {\doibase 10.1093/mnras/stw2051}
  {\bibfield  {journal} {\bibinfo  {journal} {Mon. Not. R. Astron. Soc.}\
  }\textbf {\bibinfo {volume} {463}},\ \bibinfo {pages} {712} (\bibinfo {year}
  {2016})}\BibitemShut {NoStop}%
\bibitem [{\citenamefont {{Torrealba}}\ \emph {et~al.}(2019)\citenamefont
  {{Torrealba}}, \citenamefont {{Belokurov}}, \citenamefont {{Koposov}},
  \citenamefont {{Li}}, \citenamefont {{Walker}}, \citenamefont {{Sanders}},
  \citenamefont {{Geringer-Sameth}}, \citenamefont {{Zucker}}, \citenamefont
  {{Kuehn}}, \citenamefont {{Evans}},\ and\ \citenamefont
  {{Dehnen}}}]{2019MNRAS.488.2743T}%
  \BibitemOpen
  \bibfield  {author} {\bibinfo {author} {\bibfnamefont {G.}~\bibnamefont
  {{Torrealba}}}, \bibinfo {author} {\bibfnamefont {V.}~\bibnamefont
  {{Belokurov}}}, \bibinfo {author} {\bibfnamefont {S.~E.}\ \bibnamefont
  {{Koposov}}}, \bibinfo {author} {\bibfnamefont {T.~S.}\ \bibnamefont {{Li}}},
  \bibinfo {author} {\bibfnamefont {M.~G.}\ \bibnamefont {{Walker}}}, \bibinfo
  {author} {\bibfnamefont {J.~L.}\ \bibnamefont {{Sanders}}}, \bibinfo {author}
  {\bibfnamefont {A.}~\bibnamefont {{Geringer-Sameth}}}, \bibinfo {author}
  {\bibfnamefont {D.~B.}\ \bibnamefont {{Zucker}}}, \bibinfo {author}
  {\bibfnamefont {K.}~\bibnamefont {{Kuehn}}}, \bibinfo {author} {\bibfnamefont
  {N.~W.}\ \bibnamefont {{Evans}}}, \ and\ \bibinfo {author} {\bibfnamefont
  {W.}~\bibnamefont {{Dehnen}}},\ }\bibfield  {title} {\enquote {\bibinfo
  {title} {{The hidden giant: discovery of an enormous Galactic dwarf satellite
  in Gaia DR2}},}\ }\href {\doibase 10.1093/mnras/stz1624} {\bibfield
  {journal} {\bibinfo  {journal} {Mon. Not. R. Astron. Soc.}\ }\textbf
  {\bibinfo {volume} {488}},\ \bibinfo {pages} {2743} (\bibinfo {year}
  {2019})}\BibitemShut {NoStop}%
\bibitem [{\citenamefont {{Martin}}\ \emph {et~al.}(2009)\citenamefont
  {{Martin}}, \citenamefont {{McConnachie}}, \citenamefont {{Irwin}},
  \citenamefont {{Widrow}}, \citenamefont {{Ferguson}}, \citenamefont
  {{Ibata}}, \citenamefont {{Dubinski}}, \citenamefont {{Babul}}, \citenamefont
  {{Chapman}}, \citenamefont {{Fardal}}, \citenamefont {{Lewis}}, \citenamefont
  {{Navarro}},\ and\ \citenamefont {{Rich}}}]{2009ApJ...705..758M}%
  \BibitemOpen
  \bibfield  {author} {\bibinfo {author} {\bibfnamefont {Nicolas~F.}\
  \bibnamefont {{Martin}}}, \bibinfo {author} {\bibfnamefont {Alan~W.}\
  \bibnamefont {{McConnachie}}}, \bibinfo {author} {\bibfnamefont {Mike}\
  \bibnamefont {{Irwin}}}, \bibinfo {author} {\bibfnamefont {Lawrence~M.}\
  \bibnamefont {{Widrow}}}, \bibinfo {author} {\bibfnamefont {Annette M.~N.}\
  \bibnamefont {{Ferguson}}}, \bibinfo {author} {\bibfnamefont {Rodrigo~A.}\
  \bibnamefont {{Ibata}}}, \bibinfo {author} {\bibfnamefont {John}\
  \bibnamefont {{Dubinski}}}, \bibinfo {author} {\bibfnamefont {Arif}\
  \bibnamefont {{Babul}}}, \bibinfo {author} {\bibfnamefont {Scott}\
  \bibnamefont {{Chapman}}}, \bibinfo {author} {\bibfnamefont {Mark}\
  \bibnamefont {{Fardal}}}, \bibinfo {author} {\bibfnamefont {Geraint~F.}\
  \bibnamefont {{Lewis}}}, \bibinfo {author} {\bibfnamefont {Julio}\
  \bibnamefont {{Navarro}}}, \ and\ \bibinfo {author} {\bibfnamefont
  {R.~Michael}\ \bibnamefont {{Rich}}},\ }\bibfield  {title} {\enquote
  {\bibinfo {title} {{PAndAS' CUBS: Discovery of Two New Dwarf Galaxies in the
  Surroundings of the Andromeda and Triangulum Galaxies}},}\ }\href {\doibase
  10.1088/0004-637X/705/1/758} {\bibfield  {journal} {\bibinfo  {journal}
  {\apj}\ }\textbf {\bibinfo {volume} {705}},\ \bibinfo {pages} {758} (\bibinfo
  {year} {2009})}\BibitemShut {NoStop}%
\bibitem [{\citenamefont {{Crnojevi{\'c}}}\ \emph {et~al.}(2016)\citenamefont
  {{Crnojevi{\'c}}}, \citenamefont {{Sand}}, \citenamefont {{Spekkens}},
  \citenamefont {{Caldwell}}, \citenamefont {{Guhathakurta}}, \citenamefont
  {{McLeod}}, \citenamefont {{Seth}}, \citenamefont {{Simon}}, \citenamefont
  {{Strader}},\ and\ \citenamefont {{Toloba}}}]{2016ApJ...823...19C}%
  \BibitemOpen
  \bibfield  {author} {\bibinfo {author} {\bibfnamefont {D.}~\bibnamefont
  {{Crnojevi{\'c}}}}, \bibinfo {author} {\bibfnamefont {D.~J.}\ \bibnamefont
  {{Sand}}}, \bibinfo {author} {\bibfnamefont {K.}~\bibnamefont {{Spekkens}}},
  \bibinfo {author} {\bibfnamefont {N.}~\bibnamefont {{Caldwell}}}, \bibinfo
  {author} {\bibfnamefont {P.}~\bibnamefont {{Guhathakurta}}}, \bibinfo
  {author} {\bibfnamefont {B.}~\bibnamefont {{McLeod}}}, \bibinfo {author}
  {\bibfnamefont {A.}~\bibnamefont {{Seth}}}, \bibinfo {author} {\bibfnamefont
  {J.~D.}\ \bibnamefont {{Simon}}}, \bibinfo {author} {\bibfnamefont
  {J.}~\bibnamefont {{Strader}}}, \ and\ \bibinfo {author} {\bibfnamefont
  {E.}~\bibnamefont {{Toloba}}},\ }\bibfield  {title} {\enquote {\bibinfo
  {title} {{The Extended Halo of Centaurus A: Uncovering Satellites, Streams,
  and Substructures}},}\ }\href {\doibase 10.3847/0004-637X/823/1/19}
  {\bibfield  {journal} {\bibinfo  {journal} {\apj}\ }\textbf {\bibinfo
  {volume} {823}},\ \bibinfo {eid} {19} (\bibinfo {year} {2016})}\BibitemShut
  {NoStop}%
\bibitem [{\citenamefont {{Bennet}}\ \emph {et~al.}(2017)\citenamefont
  {{Bennet}}, \citenamefont {{Sand}}, \citenamefont {{Crnojevi{\'c}}},
  \citenamefont {{Spekkens}}, \citenamefont {{Zaritsky}},\ and\ \citenamefont
  {{Karunakaran}}}]{2017ApJ...850..109B}%
  \BibitemOpen
  \bibfield  {author} {\bibinfo {author} {\bibfnamefont {P.}~\bibnamefont
  {{Bennet}}}, \bibinfo {author} {\bibfnamefont {D.~J.}\ \bibnamefont
  {{Sand}}}, \bibinfo {author} {\bibfnamefont {D.}~\bibnamefont
  {{Crnojevi{\'c}}}}, \bibinfo {author} {\bibfnamefont {K.}~\bibnamefont
  {{Spekkens}}}, \bibinfo {author} {\bibfnamefont {D.}~\bibnamefont
  {{Zaritsky}}}, \ and\ \bibinfo {author} {\bibfnamefont {A.}~\bibnamefont
  {{Karunakaran}}},\ }\bibfield  {title} {\enquote {\bibinfo {title}
  {{Discovery of Diffuse Dwarf Galaxy Candidates around M101}},}\ }\href
  {\doibase 10.3847/1538-4357/aa9180} {\bibfield  {journal} {\bibinfo
  {journal} {\apj}\ }\textbf {\bibinfo {volume} {850}},\ \bibinfo {eid} {109}
  (\bibinfo {year} {2017})}\BibitemShut {NoStop}%
\bibitem [{\citenamefont {{Mao}}\ \emph {et~al.}(2021)\citenamefont {{Mao}},
  \citenamefont {{Geha}}, \citenamefont {{Wechsler}}, \citenamefont {{Weiner}},
  \citenamefont {{Tollerud}}, \citenamefont {{Nadler}},\ and\ \citenamefont
  {{Kallivayalil}}}]{2021ApJ...907...85M}%
  \BibitemOpen
  \bibfield  {author} {\bibinfo {author} {\bibfnamefont {Yao-Yuan}\
  \bibnamefont {{Mao}}}, \bibinfo {author} {\bibfnamefont {Marla}\ \bibnamefont
  {{Geha}}}, \bibinfo {author} {\bibfnamefont {Risa~H.}\ \bibnamefont
  {{Wechsler}}}, \bibinfo {author} {\bibfnamefont {Benjamin}\ \bibnamefont
  {{Weiner}}}, \bibinfo {author} {\bibfnamefont {Erik~J.}\ \bibnamefont
  {{Tollerud}}}, \bibinfo {author} {\bibfnamefont {Ethan~O.}\ \bibnamefont
  {{Nadler}}}, \ and\ \bibinfo {author} {\bibfnamefont {Nitya}\ \bibnamefont
  {{Kallivayalil}}},\ }\bibfield  {title} {\enquote {\bibinfo {title} {{The
  SAGA Survey. II. Building a Statistical Sample of Satellite Systems around
  Milky Way-like Galaxies}},}\ }\href {\doibase 10.3847/1538-4357/abce58}
  {\bibfield  {journal} {\bibinfo  {journal} {\apj}\ }\textbf {\bibinfo
  {volume} {907}},\ \bibinfo {eid} {85} (\bibinfo {year} {2021})}\BibitemShut
  {NoStop}%
\bibitem [{\citenamefont {{Molero}}\ \emph {et~al.}(2021)\citenamefont
  {{Molero}}, \citenamefont {{Simonetti}}, \citenamefont {{Matteucci}},\ and\
  \citenamefont {{della Valle}}}]{2021MNRAS.500.1071M}%
  \BibitemOpen
  \bibfield  {author} {\bibinfo {author} {\bibfnamefont {Marta}\ \bibnamefont
  {{Molero}}}, \bibinfo {author} {\bibfnamefont {Paolo}\ \bibnamefont
  {{Simonetti}}}, \bibinfo {author} {\bibfnamefont {Francesca}\ \bibnamefont
  {{Matteucci}}}, \ and\ \bibinfo {author} {\bibfnamefont {Massimo}\
  \bibnamefont {{della Valle}}},\ }\bibfield  {title} {\enquote {\bibinfo
  {title} {{Predicted rates of merging neutron stars in galaxies}},}\ }\href
  {\doibase 10.1093/mnras/staa3340} {\bibfield  {journal} {\bibinfo  {journal}
  {Mon. Not. R. Astron. Soc.}\ }\textbf {\bibinfo {volume} {500}},\ \bibinfo
  {pages} {1071} (\bibinfo {year} {2021})}\BibitemShut {NoStop}%
\bibitem [{\citenamefont {Andrade}\ and\ \citenamefont
  {Viana}(2023)}]{Andrade:2023jul}%
  \BibitemOpen
  \bibfield  {author} {\bibinfo {author} {\bibfnamefont {Micael}\ \bibnamefont
  {Andrade}}\ and\ \bibinfo {author} {\bibfnamefont {Aion}\ \bibnamefont
  {Viana}} (\bibinfo {collaboration} {SWGO Collaboration}),\ }\bibfield
  {title} {\enquote {\bibinfo {title} {{Dark Matter searches in Dwarf Galaxies
  with the Southern Wide-field Gamma-ray Observatory}},}\ }\href {\doibase
  10.22323/1.444.1413} {\bibfield  {journal} {\bibinfo  {journal} {Proc. Sci.
  ICR2023}\ ,\ \bibinfo {pages} {1413}} (\bibinfo {year} {2023})}\BibitemShut
  {NoStop}%
\bibitem [{\citenamefont {Loewenstein}\ and\ \citenamefont
  {White}(1999)}]{Loewenstein:1999ue}%
  \BibitemOpen
  \bibfield  {author} {\bibinfo {author} {\bibfnamefont {Michael}\ \bibnamefont
  {Loewenstein}}\ and\ \bibinfo {author} {\bibfnamefont {Raymond~E.}\
  \bibnamefont {White}, \bibfnamefont {III}},\ }\bibfield  {title} {\enquote
  {\bibinfo {title} {{Prevalence and properties of dark matter in elliptical
  galaxies}},}\ }\href {\doibase 10.1086/307256} {\bibfield  {journal}
  {\bibinfo  {journal} {Astrophys. J.}\ }\textbf {\bibinfo {volume} {518}},\
  \bibinfo {pages} {50} (\bibinfo {year} {1999})}\BibitemShut {NoStop}%
\bibitem [{\citenamefont {{Buote}}\ and\ \citenamefont
  {{Humphrey}}(2012)}]{Buote:2011zk}%
  \BibitemOpen
  \bibfield  {author} {\bibinfo {author} {\bibfnamefont {David~A.}\
  \bibnamefont {{Buote}}}\ and\ \bibinfo {author} {\bibfnamefont {Philip~J.}\
  \bibnamefont {{Humphrey}}},\ }\bibfield  {title} {\enquote {\bibinfo {title}
  {{Dark Matter in Elliptical Galaxies}},}\ }in\ \href {\doibase
  10.1007/978-1-4614-0580-1_8} {\emph {\bibinfo {booktitle} {Astrophysics and
  Space Science Library}}},\ \bibinfo {series} {Astrophysics and Space Science
  Library}, Vol.\ \bibinfo {volume} {378},\ \bibinfo {editor} {edited by\
  \bibinfo {editor} {\bibfnamefont {Dong-Woo}\ \bibnamefont {{Kim}}}\ and\
  \bibinfo {editor} {\bibfnamefont {Silvia}\ \bibnamefont {{Pellegrini}}}}\
  (\bibinfo {year} {2012})\ p.\ \bibinfo {pages} {235},\ \Eprint
  {http://arxiv.org/abs/1104.0012} {arXiv:1104.0012 [astro-ph.CO]} \BibitemShut
  {NoStop}%
\bibitem [{\citenamefont {{Teyssier}}\ \emph {et~al.}(2013)\citenamefont
  {{Teyssier}}, \citenamefont {{Pontzen}}, \citenamefont {{Dubois}},\ and\
  \citenamefont {{Read}}}]{2013MNRAS.429.3068T}%
  \BibitemOpen
  \bibfield  {author} {\bibinfo {author} {\bibfnamefont {Romain}\ \bibnamefont
  {{Teyssier}}}, \bibinfo {author} {\bibfnamefont {Andrew}\ \bibnamefont
  {{Pontzen}}}, \bibinfo {author} {\bibfnamefont {Yohan}\ \bibnamefont
  {{Dubois}}}, \ and\ \bibinfo {author} {\bibfnamefont {Justin~I.}\
  \bibnamefont {{Read}}},\ }\bibfield  {title} {\enquote {\bibinfo {title}
  {{Cusp-core transformations in dwarf galaxies: observational predictions}},}\
  }\href {\doibase 10.1093/mnras/sts563} {\bibfield  {journal} {\bibinfo
  {journal} {Mon. Not. R. Astron. Soc.}\ }\textbf {\bibinfo {volume} {429}},\
  \bibinfo {pages} {3068} (\bibinfo {year} {2013})}\BibitemShut {NoStop}%
\bibitem [{\citenamefont {{de Blok}}\ and\ \citenamefont
  {{Bosma}}(2002)}]{2002A&A...385..816D}%
  \BibitemOpen
  \bibfield  {author} {\bibinfo {author} {\bibfnamefont {W.~J.~G.}\
  \bibnamefont {{de Blok}}}\ and\ \bibinfo {author} {\bibfnamefont
  {A.}~\bibnamefont {{Bosma}}},\ }\bibfield  {title} {\enquote {\bibinfo
  {title} {{High-resolution rotation curves of low surface brightness
  galaxies}},}\ }\href {\doibase 10.1051/0004-6361:20020080} {\bibfield
  {journal} {\bibinfo  {journal} {AAP}\ }\textbf {\bibinfo {volume} {385}},\
  \bibinfo {pages} {816} (\bibinfo {year} {2002})}\BibitemShut {NoStop}%
\bibitem [{\citenamefont {van~den Bosch}\ \emph {et~al.}(2000)\citenamefont
  {van~den Bosch}, \citenamefont {Robertson}, \citenamefont {Dalcanton},\ and\
  \citenamefont {de~Blok}}]{vandenBosch:1999ka}%
  \BibitemOpen
  \bibfield  {author} {\bibinfo {author} {\bibfnamefont {Frank~C.}\
  \bibnamefont {van~den Bosch}}, \bibinfo {author} {\bibfnamefont {Brant~E.}\
  \bibnamefont {Robertson}}, \bibinfo {author} {\bibfnamefont {Julianne~J.}\
  \bibnamefont {Dalcanton}}, \ and\ \bibinfo {author} {\bibfnamefont {W.~J.~G}\
  \bibnamefont {de~Blok}},\ }\bibfield  {title} {\enquote {\bibinfo {title}
  {{Constraints on the structure of dark matter halos from the rotation curves
  of low surface brightness galaxies}},}\ }\href {\doibase 10.1086/301315}
  {\bibfield  {journal} {\bibinfo  {journal} {Astron. J.}\ }\textbf {\bibinfo
  {volume} {119}},\ \bibinfo {pages} {1579} (\bibinfo {year}
  {2000})}\BibitemShut {NoStop}%
\bibitem [{\citenamefont {{van den Bosch}}\ and\ \citenamefont
  {{Swaters}}(2001)}]{2001MNRAS.325.1017V}%
  \BibitemOpen
  \bibfield  {author} {\bibinfo {author} {\bibfnamefont {Frank~C.}\
  \bibnamefont {{van den Bosch}}}\ and\ \bibinfo {author} {\bibfnamefont
  {Rob~A.}\ \bibnamefont {{Swaters}}},\ }\bibfield  {title} {\enquote {\bibinfo
  {title} {{Dwarf galaxy rotation curves and the core problem of dark matter
  haloes}},}\ }\href {\doibase 10.1046/j.1365-8711.2001.04456.x} {\bibfield
  {journal} {\bibinfo  {journal} {Mon. Not. R. Astron. Soc.}\ }\textbf
  {\bibinfo {volume} {325}},\ \bibinfo {pages} {1017} (\bibinfo {year}
  {2001})}\BibitemShut {NoStop}%
\bibitem [{\citenamefont {Simon}\ \emph {et~al.}(2004)\citenamefont {Simon},
  \citenamefont {Bolatto}, \citenamefont {Leroy},\ and\ \citenamefont
  {Blitz}}]{Simon:2003tf}%
  \BibitemOpen
  \bibfield  {author} {\bibinfo {author} {\bibfnamefont {Joshua~D.}\
  \bibnamefont {Simon}}, \bibinfo {author} {\bibfnamefont {Alberto~D.}\
  \bibnamefont {Bolatto}}, \bibinfo {author} {\bibfnamefont {Adam}\
  \bibnamefont {Leroy}}, \ and\ \bibinfo {author} {\bibfnamefont {Leo}\
  \bibnamefont {Blitz}},\ }\bibfield  {title} {\enquote {\bibinfo {title}
  {{Dark matter in dwarf galaxies: Latest density profile results}},}\ }\href
  {\doibase 10.48550/arXiv.astro-ph/0310193} {\bibfield  {journal} {\bibinfo
  {journal} {ASP Conf. Ser.}\ }\textbf {\bibinfo {volume} {327}},\ \bibinfo
  {pages} {18} (\bibinfo {year} {2004})}\BibitemShut {NoStop}%
\bibitem [{\citenamefont {Simon}\ \emph {et~al.}(2003)\citenamefont {Simon},
  \citenamefont {Bolatto}, \citenamefont {Leroy},\ and\ \citenamefont
  {Blitz}}]{Simon:2003xu}%
  \BibitemOpen
  \bibfield  {author} {\bibinfo {author} {\bibfnamefont {Joshua~D.}\
  \bibnamefont {Simon}}, \bibinfo {author} {\bibfnamefont {Alberto~D.}\
  \bibnamefont {Bolatto}}, \bibinfo {author} {\bibfnamefont {Adam}\
  \bibnamefont {Leroy}}, \ and\ \bibinfo {author} {\bibfnamefont {Leo}\
  \bibnamefont {Blitz}},\ }\bibfield  {title} {\enquote {\bibinfo {title}
  {{High-resolution measurements of the dark matter halo of ngc 2976: evidence
  for a shallow density profile}},}\ }\href {\doibase 10.1086/378200}
  {\bibfield  {journal} {\bibinfo  {journal} {Astrophys. J.}\ }\textbf
  {\bibinfo {volume} {596}},\ \bibinfo {pages} {957} (\bibinfo {year}
  {2003})}\BibitemShut {NoStop}%
\bibitem [{\citenamefont {Pacucci}\ \emph {et~al.}(2023)\citenamefont
  {Pacucci}, \citenamefont {Ni},\ and\ \citenamefont {Loeb}}]{Pacucci:2023eaf}%
  \BibitemOpen
  \bibfield  {author} {\bibinfo {author} {\bibfnamefont {Fabio}\ \bibnamefont
  {Pacucci}}, \bibinfo {author} {\bibfnamefont {Yueying}\ \bibnamefont {Ni}}, \
  and\ \bibinfo {author} {\bibfnamefont {Abraham}\ \bibnamefont {Loeb}},\
  }\bibfield  {title} {\enquote {\bibinfo {title} {{Extreme Tidal Stripping May
  Explain the Overmassive Black Hole in Leo I: A Proof of Concept}},}\ }\href
  {\doibase 10.3847/2041-8213/acff5e} {\bibfield  {journal} {\bibinfo
  {journal} {Astrophys. J. Lett.}\ }\textbf {\bibinfo {volume} {956}},\
  \bibinfo {pages} {L37} (\bibinfo {year} {2023})}\BibitemShut {NoStop}%
\end{thebibliography}%

\end{document}